\documentclass[12pt]{article}
\usepackage{amsmath,epsfig,euscript,bm,url}

\def\pslash{\rlap{\hspace{0.02cm}/}{p}}
\def\kslash{\rlap{\hspace{0.02cm}/}{k}}
\def\lslash{\rlap{\hspace{-0.02cm}/}{l}}
\def\qslash{\rlap{\hspace{-0.01cm}/}{q}}
\def\vslash{\rlap{\hspace{0.02cm}/}{v}}
\def\mproton{M}
\newcommand{\imagespace}{1.25em}

\begin{document}

\begin{titlepage}

\begin{flushright}
FERMILAB-PUB-18-675-T\\
WSU-HEP-1808\\
December 12, 2018
\end{flushright}

\vspace{0.7cm}
\begin{center}
\Large\bf\boldmath
Elements of QED-NRQED Effective Field Theory:\\ II. Matching of Contact Interactions 
\unboldmath
\end{center}

\vspace{0.8cm}
\begin{center}
{\sc Steven P. Dye$^{(a)}$, Matthew Gonderinger$^{(a)}$, Gil Paz$^{(a,b)}$}\\
\vspace{0.4cm}
{\it 
$^{(a)}$ Department of Physics and Astronomy, \\
Wayne State University, Detroit, Michigan 48201, USA 
}\\
\vspace{0.3cm}
{\it 
$^{(b)}$ Theoretical Physics Department, \\Fermilab, P.O. Box 500, Batavia, Illinois 60510, USA
}

\end{center}
\vspace{0.7cm}
\begin{abstract}
  \vspace{0.2cm}
  \noindent
In 2010 the first extraction of the proton charge radius from muonic hydrogen was found be five standard deviations away from the regular hydrogen value. Eight years later, this proton radius puzzle is still unresolved.  One of the most promising avenues to resolve the puzzle is by a muon-proton scattering experiment called MUSE. The typical momenta of the muons in this experiment are of the order of the muon mass. In this energy regime the muons are relativistic but the protons are non-relativistic. The interaction between them can be  described by QED-NRQED effective field theory. In a previous paper we have shown how QED-NRQED reproduces Rosenbluth scattering up to $1/M^2$, where $M$ is the proton mass,  and relativistic scattering off a static potential at ${\cal O}(Z^2\alpha^2)$ and leading power in $M$. In this paper we determine the Wilson coefficients of the four-fermion contact interactions at ${\cal O}(Z^2\alpha^2)$ and power $1/M^2$. Surprisingly, we find that the coefficient of the spin-independent interaction vanishes, implying that MUSE will be sensitive mostly to the proton charge radius and not spin-independent two-photon exchange effects.    
\end{abstract}
\vfil

\end{titlepage}
\section{Introduction}
The response of an on-shell proton to a one-photon electromagnetic probe can be parametrized in terms of two non-perturbative form factors, the so-called ``Dirac" ($F_1(q^2)$) and ``Pauli" ($F_2(q^2)$) form factors, see (\ref{FF}) below. Alternatively, one can define a different linear combination as the ``electric" ($G_E=F_1+q^2F_2/4M^2$, where $M$ is the proton mass) and ``magnetic" ($G_M=F_1+F_2$) form factors. The slope of $G_E$ at $q^2=0$  \emph{defines} the proton charge radius (squared), as $\left(r_E^p\right)^2 =6\,dG^{p}_E(q^2)/dq^2|_{q^2=0}$. The two main methods to extract $r_E^p$ are lepton-proton scattering and spectroscopy of lepton-proton bound states, i.e. regular or muonic hydrogen. Until 2010, only values from electron-proton scattering and regular hydrogen were available. In 2010, the first extraction of $r_E^p$ from muonic hydrogen was reported as $r_E^p =0.84184(67)$ fm \cite{Pohl:2010zza}. Surprisingly, it was five standard deviations lower than the regular hydrogen value of $r_E^p =0.8768(69)$ fm \cite{Mohr:2008fa}. The most recent update from muonic hydrogen is $r_E^p =0.84087(39)$ fm \cite{Antognini:1900ns}, while the 2014 CODATA value is $r_E^p =0.8751(61)$ fm \cite{Mohr:2015ccw}. This discrepancy is known as the ``proton radius puzzle". 

The most exciting interpretation of the puzzle is as a new interaction that distinguishes electrons and muons, see, e.g., the references in \cite{Epstein:2014zua,Carlson:2015jba}. In such explanations the crucial questions are what assumptions one is making about this new interaction, in particular how does it couple to hadrons besides protons. At the same time, one would like to test more mundane explanations such as experimental or theoretical issues in the extractions of $r_E^p$ from spectroscopy or scattering using both electrons and muons\footnote{We focus on hydrogen. More theoretical and experimental work was also done for deuterium and helium.}. We briefly review those.

 {\bf Regular hydrogen spectroscopy:}  No issues were identified with the theoretical input. On the experimental side, new measurements were published in the last two years with error bars comparable to the 2014 CODATA value. In 2017 a group in Garching has published the value $r_E^p = 0.8335(95)$ fm  from $2S-4P$ transition \cite{Beyer:2017}. In 2018 a group in Paris has published the value $r_E^p =0.877(13)$ fm from $1S-3S$ transition \cite{Fleurbaey:2018fih}. More results are expected in the very near future \cite{PRP2018}.    
 
{\bf Muonic hydrogen spectroscopy:} No issues were raised concerning the muonic hydrogen experimental measurement. It should be noted that the extraction of $r_E^p$ from muonic hydrogen spectroscopy is based on the work of one group.  There are no plans by other groups to repeat the measurement. On the theoretical side there was much discussion in the literature. It is reflected in the different theoretical formula used in \cite{Pohl:2010zza} and the 2013 update \cite{Antognini:1900ns}. Due to its precision, the muonic hydrogen result involves a more complicated hadronic input, beyond a one-photon probe of the proton structure. In particular, as emphasized in \cite{Hill:2011wy}, two-photon effects are a potential source of uncertainty. The imaginary part of the two-photon exchange amplitude is related to experimental data: form factors and structure functions, see section \ref{subsec:proton} below. The amplitude cannot be reconstructed from its imaginary part and the knowledge of a subtraction function $W_1(0,Q^2)$, where $Q^2=-q^2$, is required, see (\ref{W1disp}) below. The subtraction function $W_1(0,Q^2)$ is not known exactly. Its small $Q^2$ expression is calculable using Non-Relativistic QED (NRQED) \cite{Hill:2011wy}. Its large $Q^2$ expression can be calculated using the operator product expansion. The spin-0 contribution was calculated in \cite{Collins:1978hi} and corrected in \cite{Hill:2016bjv}. The spin-2 contribution was calculated in \cite{Hill:2016bjv}. In \cite{Hill:2016bjv} the two limits were interpolated to give an estimate for the contribution of $W_1(0,Q^2)$ to two-photon exchange effects. The uncertainty on the interpolation is larger than in \cite{Antognini:1900ns}, but it is too small to explain the discrepancy. The estimate in \cite{Hill:2016bjv} is consistent with the literature \cite{Pachucki:1999zza, Martynenko:2005rc, Nevado:2007dd, Carlson:2011zd, Birse:2012eb, Gorchtein:2013yga, Alarcon:2013cba, Peset:2014jxa}. On the other hand, \cite{Miller:2012ne} finds a much larger uncertainty. Ultimately one would like to probe the muon-proton two-photon exchange effects by using a different method such as muon-proton scattering.   

{\bf Electron-proton scattering:} Extractions of $r_E^p$ from the electron-proton cross section data or even  the form factor itself require an extrapolation to $q^2=0$. Since $G_E$'s functional form is not known, such an extrapolation is not simple. In \cite{Hill:2010yb} it was first suggested to utilize the $z$ expansion, used before for meson form factors, to perform the extrapolation.  The $z$ expansion is based on the known analytic properties of the form factor. The values obtained using $z$-expansion analyses \cite{Hill:2010yb, Lee:2015jqa} generally disfavor the muonic hydrogen result\footnote{Some other $z$-expansion based studies do not bound the coefficients of the $z$ expansion \cite{Lorenz:2014vha, Lorenz:2014yda, Griffioen:2015hta}. These may result in values that are lower than \cite{Hill:2010yb,Lee:2015jqa}. See \cite{Hill:2010yb} for a discussion of the bounding of the coefficients.}. More recently, lattice extractions of $r_E^p$ have used the $z$ expansion, see e.g.  \cite{Alexandrou:2017ypw,Jang:2018lup, Shintani:2018ozy}, although currently the errors are typically too large to distinguish between the two values of $r_E^p$.  Other recent extractions use dipole \cite{Higinbotham:2015rja},  polynomial \cite{Griffioen:2015hta}, continued fraction \cite{Griffioen:2015hta}, modified $z$ expansion \cite{Horbatsch:2015qda}, or other parameterizations of the form factors, as well as include inputs from chiral EFT \cite{Horbatsch:2016ilr,Alarcon:2018zbz}. Most of these \cite{Higinbotham:2015rja,Griffioen:2015hta,Alarcon:2018zbz} favor the muonic hydrogen result. For pre-2010 extractions see \cite{PDG:2014}.

On the experimental side a new low-$Q^2$ electron-proton scattering experiment called ``PRad" \cite{Gasparian:2017cgp} was recently performed at Jefferson Lab and its results are expected in the near future.  

{\bf Muon-proton scattering:} This is the least studied method to extract $r_E^p$. To address that, a new muon-proton scattering experiment called MUSE is being built at the Paul Scherrer Institute \cite{Gilman:2017hdr}. It will start taking data in 2019.  It is the first muon scattering measurement with the required precision to address the proton radius puzzle \cite{PRP2018}. 

In making predictions for MUSE, a phenomenological approach was taken in \cite{Tomalak:2014dja, Tomalak:2015hva, Tomalak:2017owk, Tomalak:2018jak}. One can also use effective field theory (EFT) methods. 
In muonic hydrogen the muon's typical momentum is $m\alpha\sim1$ MeV, and both the muon and the proton can be treated non-relativistically.  For MUSE the muon's typical momentum is about the muon mass $m\sim100$ MeV, and the muon must be treated relativistically, while the proton can be treated non-relativistically. An EFT for such kinematics\footnote{The dynamical degrees of freedom of this theory are proton, muon, and photon. The pion is not included as a dynamical degree of freedom. This is different from an earlier EFT applicable to the MUSE kinematics considered in \cite{Pineda:2002as, Pineda:2004mx} that contains very similar operators. See section \ref{sec:General}  for a more detailed discussion.}, called QED-NRQED, was suggested in \cite{Hill:2012rh}.

In a previous paper we  studied some aspects of this EFT \cite{Dye:2016uep}. Denoting by $m (M)$ the muon (proton) mass and using  $Z=1$ for a proton, we showed that one-photon exchange ${\cal O} (Z\alpha)$ QED-NRQED scattering at power $1/M^2$ reproduces Rosenbluth scattering \cite{Rosenbluth:1950yq}, and the two-photon exchange ${\cal O} (Z^2\alpha^2)$ QED-NRQED scattering at leading power reproduces the scattering of a relativistic fermion off a static potential \cite{Dalitz:1951ah, Itzykson:1980rh}.  

Two photon exchange effects sensitive to the proton structure start at ${\cal O} (Z^2\alpha^2)$ and power $1/M^2$. For QED-NRQED they appear as two four-fermion operators, one spin-independent and one spin-dependent, see (\ref{contactQN}) below.  This paper's goal is to determine their Wilson coefficients in terms of the proton's hadronic tensor, i.e. performing a matching onto QED-NRQED.    

The paper is structured as follows. In section \ref{sec:General} we present an overview of the general features of the matching calculation. In section \ref{sec:QN} we perform the effective field theory calculation. In section \ref{sec:Full} we perform the full theory calculation for the toy example of a non-relativistic point particle and for the proton. In section \ref{sec:Extraction} we extract the matching coefficients for both full theories. We present our conclusions in  section \ref{sec:conclusions}. The appendices describe the matching in Coulomb gauge, properties of the hadronic tensor, and list the NRQED Feynman rules.

\section{General features of the matching calculation}\label{sec:General}
\subsection{Lagrangian}
We consider the interaction between a relativistic spin-half field (``lepton") of mass $m$, denoted by $\ell$, and a non-relativistic spin-half field (``non-relativistic proton") of mass $M$, denoted by $\psi$. For the calculation of the Wilson coefficients of the leading power four-fermion operators from the proton's hadronic tensor, we will need the interactions at ${\cal O}(Z^2\alpha^2)$ and power $1/M^2$, where $Z=1$ for a proton. The QED-NRQED Lagrangian contains three parts that are relevant to the calculation. 

The first part is the NRQED Lagrangian\footnote{At this power there are also interactions of four non-relativistic spin-half fields, see \cite{Dye:2016uep}. We will not need them in this paper.}, up to $1/M^2$:
\begin{equation}\label{Lagrangian2}
{\cal L}_\psi = \psi^\dagger\left\{ i D_t +c_2\dfrac{\bm {D}^2}{2M}+c_Fe\dfrac{\bm {\sigma\cdot B}}{2M}+c_De\dfrac{[\bm{\nabla\cdot E}]}{8M^2}+ic_Se\dfrac{\bm{\sigma}\cdot\left(\bm{D\times E}-\bm{E\times D}\right)}{8M^2}\right\}\psi +\cdots,
\end{equation}
where $D_t=\partial/\partial t+ieZA^0$,  $\bm D=\bm\nabla-ieZ\bm A$, $\bm\sigma$ are the Pauli matrices, and $e$  is the electromagnetic coupling constant\footnote{We follow the conventions of \cite{Kinoshita:1995mt}, although in that paper the NRQED Lagrangian describes an electron. In other words, as in  \cite{Dye:2016uep}, we take $e$ to be positive.}. These are the components of $D_\mu=\partial_\mu+ieZA_\mu$. The notation $[\bm{\nabla\cdot E}]$ denotes that the derivative is acting only on $\bm E$ and not on $\psi$. For a review see \cite{Paz:2015uga}. The (hidden) Lorentz invariance of the Lagrangian, also known as reparameterization invariance, implies that $c_2=1$ \cite{Luke:1992cs, Manohar:1997qy, Brambilla:2003nt, Heinonen:2012km, Hill:2012rh}. The other Wilson coefficients can be related to the proton electromagnetic form factors defined by
\begin{equation}\label{FF}
\langle p(p')|J_\mu^{\rm em}|p(p)\rangle=\bar u(p')
\left[\gamma_\mu F_1(q^2)+\frac{i\sigma_{\mu\nu}}{2M}F_2(q^2)q^\nu\right]u(p)\,,
\end{equation}
via $Z=F_1(0)$, $c_F=F_1(0)+F_2(0)$,  $c_D=F_1(0)+2F_2(0)+8M^2F_1'(0)$, where $F_1^\prime=dF_1(q^2)/dq^2$, and $c_S=2c_F-F_1(0)=F_1(0)+2F_2(0)$. The latter can also be determined by the hidden Lorentz invariance of the Lagrangian \cite{Manohar:1997qy, Brambilla:2003nt,Heinonen:2012km, Hill:2012rh}. The NRQED Feynman rules are listed in appendix \ref{app:FR}. Including radiative corrections introduces scale dependence in $c_i$, but such effects will not be considered in this paper. See \cite{Manohar:1997qy,Pineda:2004mx,Hill:2012rh,Lee:2015jqa} and references within for a discussion of this issue and form factor definitions in the presence of radiative corrections.

The second part is the usual QED Lagrangian that describes the lepton's interaction with the electromagnetic field: \begin{equation}\label{LagrangianDirac}
{\cal L}_\ell=\bar\ell\, \gamma^\mu\,i\left(\partial_\mu+ieQ_\ell A_\mu \right)\ell-m\bar\ell\ell,
\end{equation}
where $Q_\ell=-1$ for a muon or an electron\footnote{We do not include $1/M^2$ operators of  \cite{Hill:2012rh}, since they have Wilson coefficients that start at ${\cal O}(\alpha)$. At the lowest order in $\alpha$ these operators lead to terms of order ${\cal O}(Z\alpha^2)$ but not ${\cal O}(Z^2\alpha^2)$ which are relevant to the QED-NRQED contact interactions.}.  

The third part is the QED-NRQED contact interactions. At $1/M^2$ we have two possible contact interactions,  
\begin{equation}\label{contactQN} 
{\cal L}_{\psi\ell}=\dfrac{b_1}{M^2}\psi^\dagger\psi\,\bar \ell\gamma^0\ell+\dfrac{b_2}{M^2}\psi^\dagger\sigma^i\psi\,\bar \ell\gamma^i\gamma^5\ell +{\cal O}\left(1/M^3\right),
\end{equation}
where our notation follows that of \cite{Hill:2012rh}. The main goal of this paper is to express the Wilson coefficients $b_1$ and $b_2$ in terms of the components of the hadronic tensor. In the matching we will encounter two other operators that are explicitly suppressed by $m/M^3$. These are $b_3\,m\psi^\dagger\psi\,\bar \ell\ell/M^3$ and $2b_6\,m\psi^\dagger\sigma^i\psi\bar \ell\left(\frac{i}2\epsilon^{ijk}\gamma^j\gamma^k\right)\ell/M^3$, in the notation of \cite{Hill:2012rh}. The other $1/M^3$ contact interactions contain space-like covariant derivatives.  Since we will set the external three-momenta to zero in the matching, we will not encounter them. The $b_i$'s are determined by matching the full hadronic tensor onto QED-NRQED. As a result, the Wilson coefficients $b_i$ depend on a scale $\Lambda\sim M$. For the MUSE experiment a scale of around the muon mass $\sim100$ MeV might be more appropriate. We will not consider running effects between the two scales in this paper but such effects are expected to be small since $Q_l^2Z^2\alpha^2\log(m/M)$ is not large for $Z=1$.  

The operators in (\ref{Lagrangian2}) and (\ref{contactQN}) were considered before in \cite{Pineda:2002as} as part of a larger Heavy Baryon Effective Theory (HBET) that contains also the pion and $\Delta$ as degrees of freedom and the interaction between them and the proton, lepton (muon or electron) and photon. Since we wish to describe only the scattering of a muon and a proton, and not other interactions, e.g. a proton and a pion, we can limit ourselves to (\ref{Lagrangian2}), (\ref{LagrangianDirac}), and (\ref{contactQN}). The effects of the strong interaction are encoded in the non-perturbative QED-NRQED Wilson coefficients $c_i$ and $b_i$, which in turn are determined by the proton form factors and the hadronic tensor, respectively. In this paper we take them as non-perturbative inputs. In particular, we determine implicit expressions for $b_i$ in terms of the components of the hadronic tensor, see equations (\ref{b1extraction}) and (\ref{b2extraction}) below. It could be interesting to consider explicit expressions for the components of the hadronic tensor, based on theoretical calculations, see e.g. \cite{Peset:2014jxa}, or from scattering data, as was done in \cite{Carlson:2011zd}, but this goes beyond the scope of this paper.

\subsection{Infrared and ultraviolet singularities}
We calculate the amplitude for forward off-shell $\ell+p\to\ell+p$ scattering at ${\cal O}(Z^2\alpha^2)$ and power $1/M^2$ in both the full and the effective theories. The former is expressed in terms of the hadronic tensor defined below, and the latter in terms of the Wilson coefficients of the effective theory. Both amplitudes are IR  singular. We regulate these by using a fictitious photon ``mass" denoted by $\lambda$. In terms of powers of $\lambda$ we will find singularities\footnote{Terms that scale like $1/\lambda^2$ cancel in the sum of direct and crossed diagrams and do not appear in the total amplitude.} that scale like $1/\lambda^3$, $1/\lambda^2$, $1/\lambda$, and $\log \lambda$.

The lepton mass $m$ is also an IR quantity. Thus we find both in the full and effective theory terms that diverge in the limit $m\to0$. In terms of powers of $m$ we will find singularities that scale like $1/m^2$, $1/m$, and $\log m$. The IR terms regulated by $\lambda$, $m$, or both, must cancel in the matching. This serves as a non-trivial check of the calculation. 

We assume $m,\lambda\ll M$, so we will expand the singular terms in powers of $m/M$ and $\lambda/M$. This allows to simplify the matching calculation. This obviously holds for the muon-proton case where $m/M\sim0.1$.

The effective field theory is valid up to a cutoff scale $\Lambda\sim M$. We use a hard momentum cutoff to regularize the UV singularities of the QED-NRQED integrals.  In the matching we also need to expand the form factors in powers of $q^2$. This can introduce UV divergent terms in the full theory calculation. In practice this is not a problem for the matching calculation, since such term are regularized when using the full functional form of the form factors. 
\subsection{Gauge Invariance}
Since the photon propagator in Coulomb gauge is different for space and time components, it is often used for NRQED calculations, see e.g. \cite{Kinoshita:1995mt}. In the following we perform the matching both in Feynman and Coulomb gauges. Both the full and effective field theory amplitudes are off-shell and as a result each amplitude need not be gauge invariant by itself.  As we will see, both the full and effective theory calculations are different for each gauge, but the Wilson coefficients $b_1$ and $b_2$ are the same. 

Interestingly,  when we take the non-relativistic limit for the lepton, the effective field theory amplitude is the same in both gauges. In this limit one adds $\psi^\dagger\psi\,\bar \ell\gamma^0\ell$ and $\psi^\dagger\psi\,\bar \ell\ell$ to form the spin-independent amplitude and $\psi^\dagger\sigma^i\psi\,\bar \ell\gamma^i\gamma^5\ell$ and $\psi^\dagger\sigma^i\psi\bar \ell\left(\frac{i}2\epsilon^{ijk}\gamma^j\gamma^k\right)\ell$ to form the spin-dependent amplitude. 

The difference between the gauges arises even for the simpler case of non-relativistic point particle interacting with a relativistic point particle. As we show in appendix \ref{app:GI}, the amplitude is the same for Feynman and general covariant  gauge, but different in Coulomb gauge. Taking the non-relativistic limit for both particles results in a gauge invariant amplitude. In a nutshell, the reason is that for $p=(m,\vec 0)$ and $\chi$  a non-relativistic spinor,  $(\pslash-m)(\chi\,\, 0)^{T}=(m\gamma^0-m)(\chi\,\,  0)^T=0$ making the non-relativistic spinor effectively on-shell in this limit. The details are given in appendix \ref{app:GI}.    

In the following we present results in Feynman gauge. The results in Coulomb gauge  are presented in appendix \ref{app:CG}.

\section{Effective Field Theory Calculation}\label{sec:QN} 
\begin{figure}
\begin{center}
\includegraphics[scale=0.6]{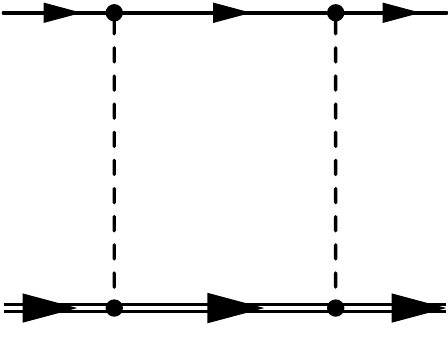} \vspace{\imagespace}
\includegraphics[scale=0.6]{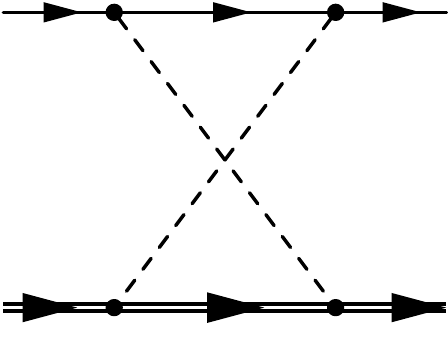}\quad
\includegraphics[scale=0.6]{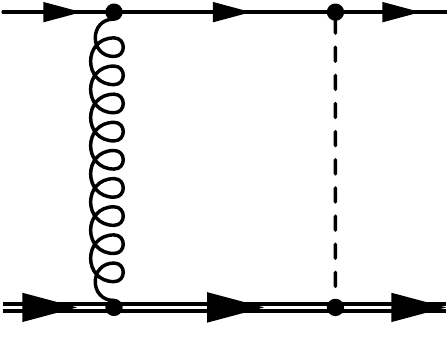}\quad
\includegraphics[scale=0.6]{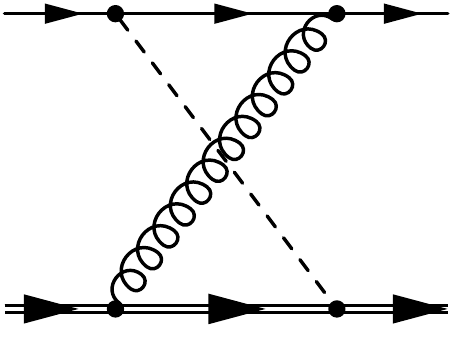}\quad
\includegraphics[scale=0.6]{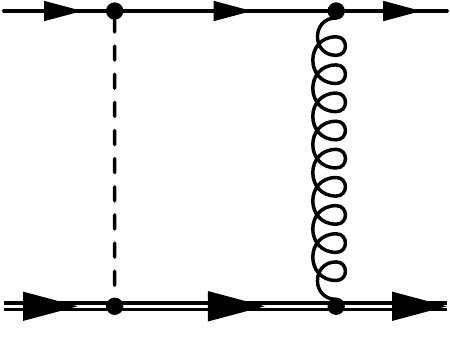} \vspace{\imagespace}
\includegraphics[scale=0.6]{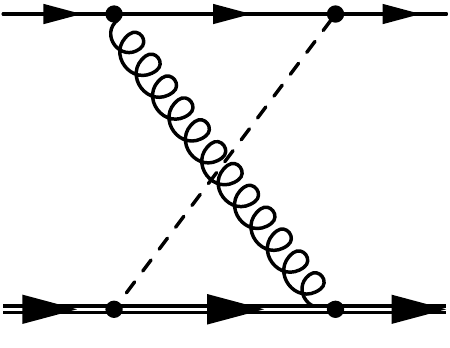}\quad
\includegraphics[scale=0.6]{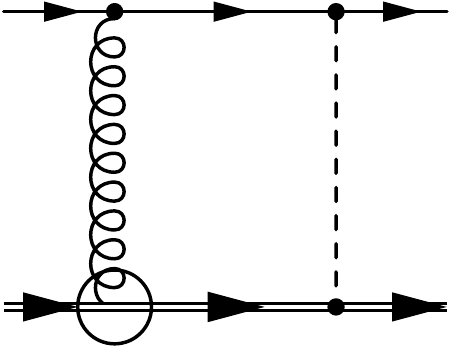}\quad
\includegraphics[scale=0.6]{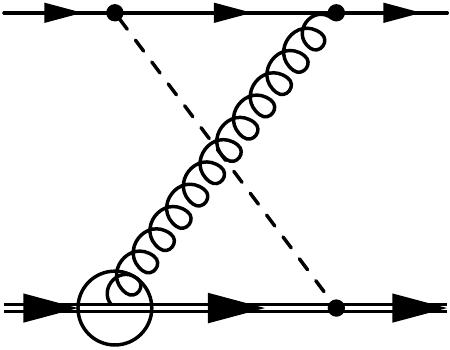}\quad
\includegraphics[scale=0.6]{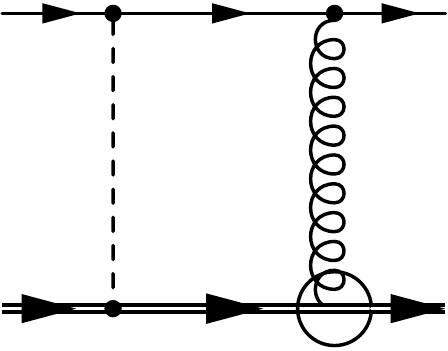}\quad
\includegraphics[scale=0.6]{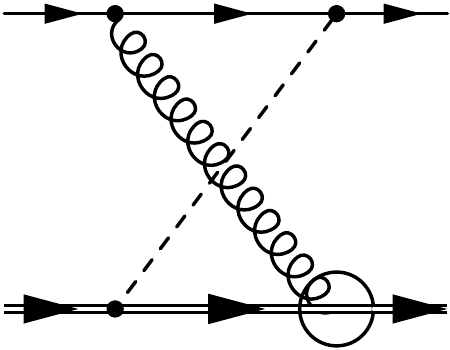}\quad \vspace{\imagespace}
\includegraphics[scale=0.6]{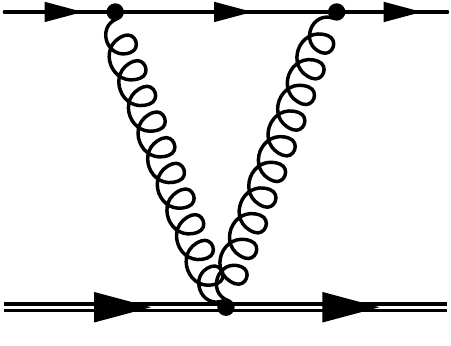}\quad
\includegraphics[scale=0.6]{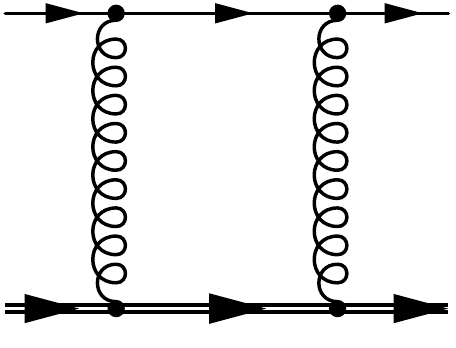}\quad
\includegraphics[scale=0.6]{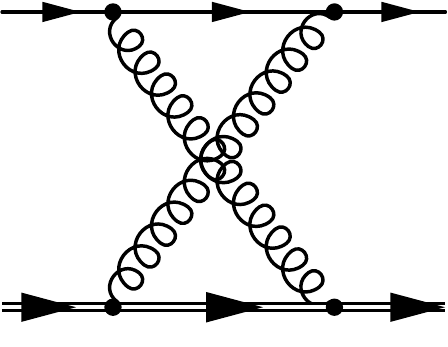}\quad
\includegraphics[scale=0.6]{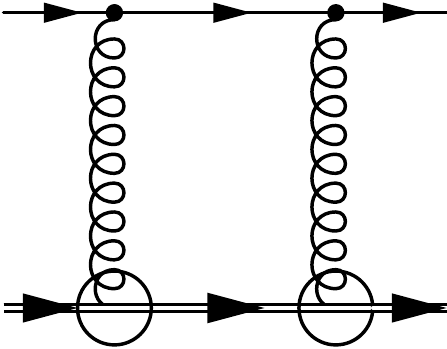}\quad
\includegraphics[scale=0.6]{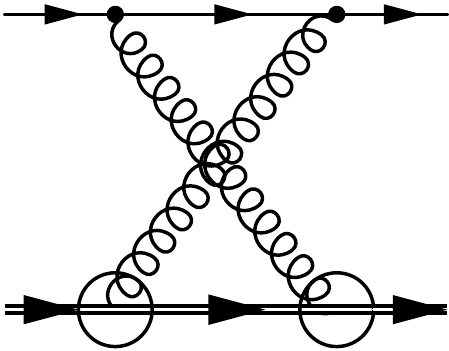}\quad \vspace{\imagespace}
\includegraphics[scale=0.6]{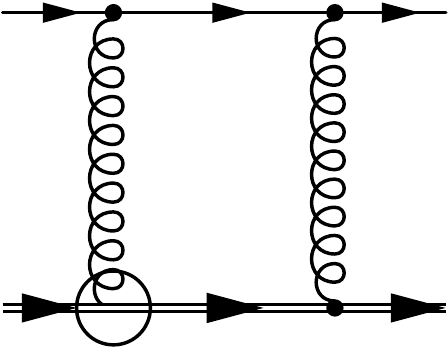}\quad
\includegraphics[scale=0.6]{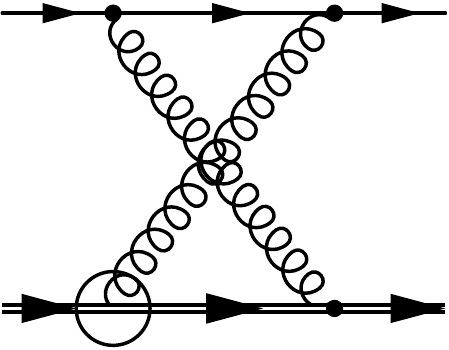}\quad
\includegraphics[scale=0.6]{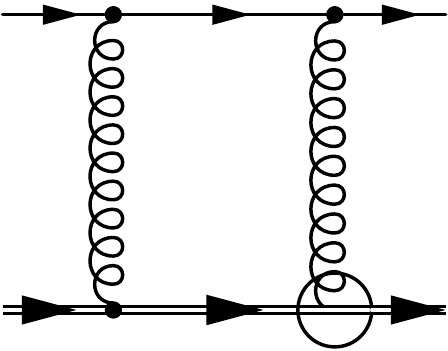}\quad
\includegraphics[scale=0.6]{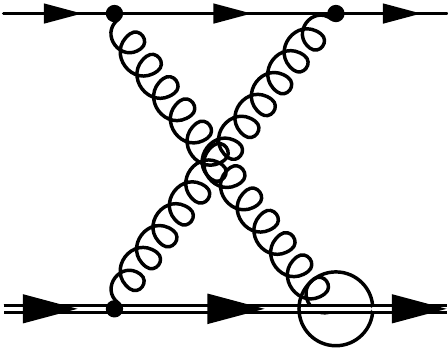}\quad
\includegraphics[scale=0.6]{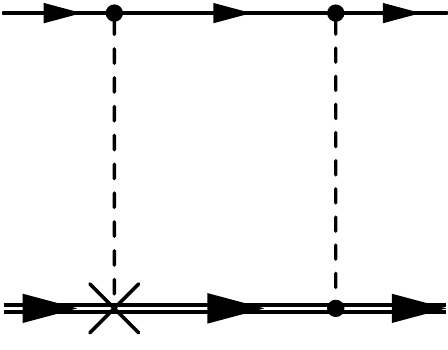}\quad \vspace{\imagespace}
\includegraphics[scale=0.6]{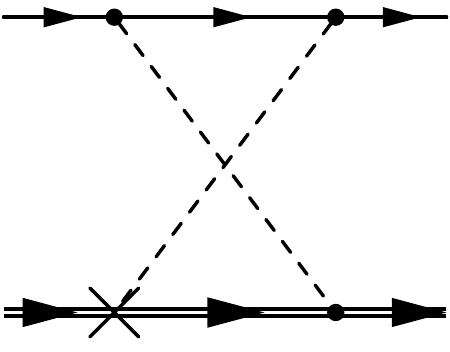}\quad
\includegraphics[scale=0.6]{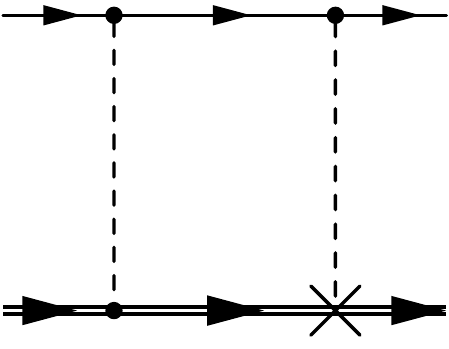}\quad
\includegraphics[scale=0.6]{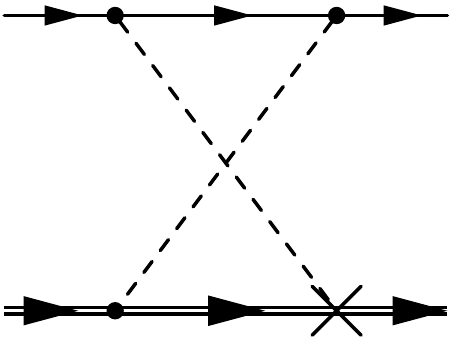}\quad
\includegraphics[scale=0.6]{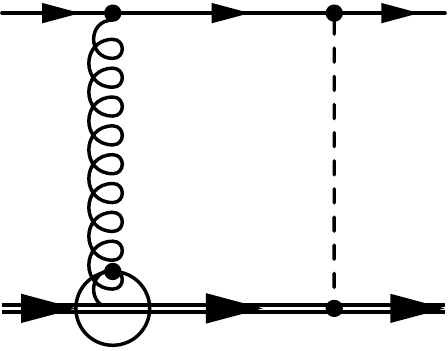}\quad
\includegraphics[scale=0.6]{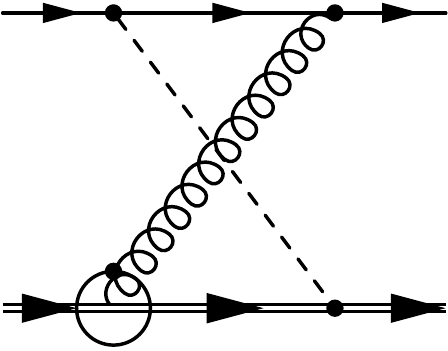}\quad \vspace{\imagespace}
\includegraphics[scale=0.6]{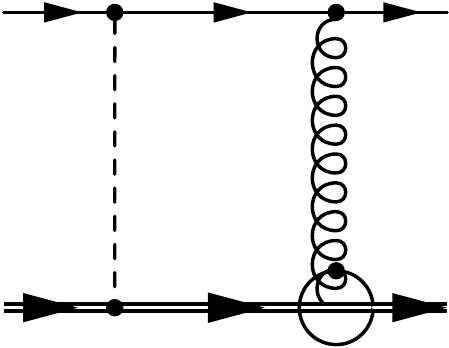}\quad
\includegraphics[scale=0.6]{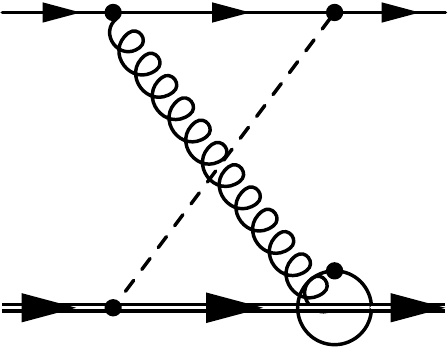}\quad
\includegraphics[scale=0.6]{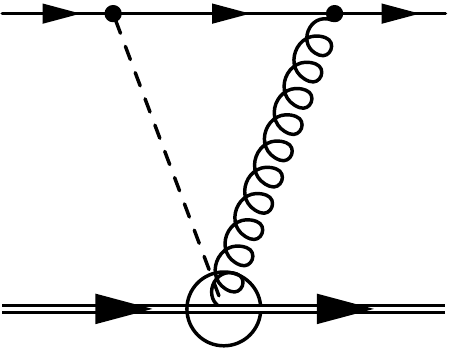}\quad
\includegraphics[scale=0.6]{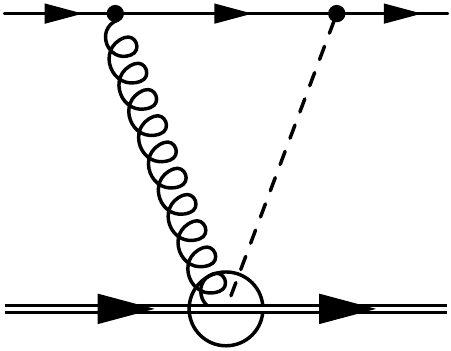}\quad
\includegraphics[scale=0.6]{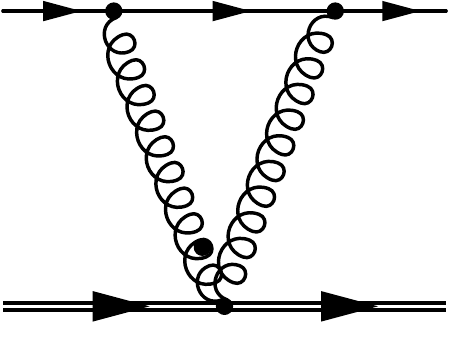}\quad \vspace{\imagespace}
\includegraphics[scale=0.6]{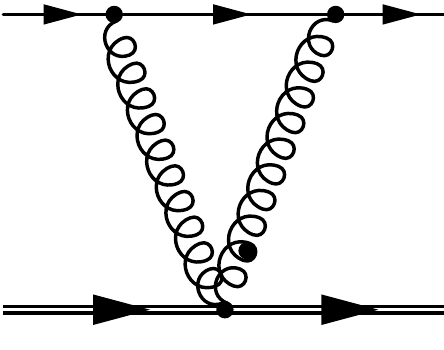}\quad
\includegraphics[scale=0.6]{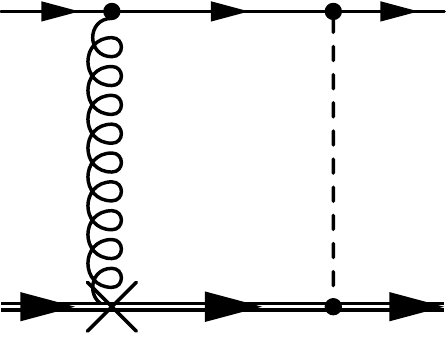}\quad
\includegraphics[scale=0.6]{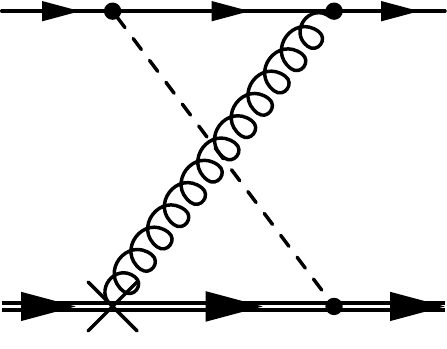}\quad
\includegraphics[scale=0.6]{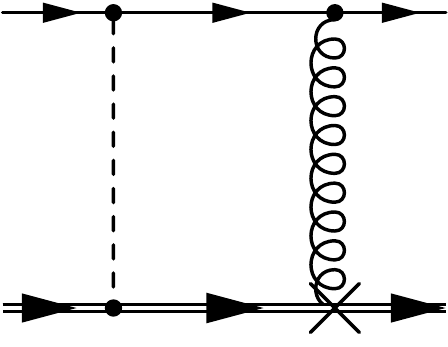}\quad
\includegraphics[scale=0.6]{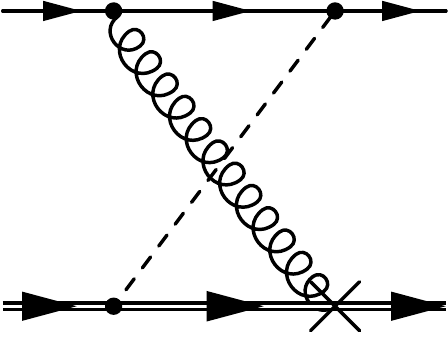}\quad
\caption{\label{QN_Diagrams} QED-NRQED two-photon exchange Feynman diagrams that contribute to forward lepton-proton scattering at ${\cal O}(Z^2\alpha^2)$ up to power $1/M^2$. The double line denotes the NRQED field. The dashed (curly) line represents Coulomb (transverse) photon. The dot, circle, cross, circle-dot, and dotted curly line symbols are defined in appendix \ref{app:FR}.  
}
\end{center}
\end{figure}

We calculate the amplitude for forward off-shell $\ell+p\to\ell+p$ scattering at ${\cal O}(Z^2\alpha^2)$ and power $1/M^2$ in QED-NRQED effective theory using Feynman gauge. The QED-NRQED calculation in Coulomb gauge is presented in appendix \ref{app:CG}. The interactions of the relativistic fermion ($\ell$) with the photon arise from (\ref{LagrangianDirac}). The interactions of the non-relativistic fermion ($p$) with the photon arise from (\ref{Lagrangian2}) and the Feynman rules are given in appendix \ref{app:FR}. Combining the two we find the diagrams shown in figure \ref{QN_Diagrams} that contribute at power $1/M^2$. Since $b_1$ and $b_2$ get no contribution at ${\cal O}(Z\alpha)$ \cite{Dye:2016uep}, there are no loop diagrams that involve (\ref{contactQN}) at ${\cal O}(Z^2\alpha^2)$ and power $1/M^2$. The tree level diagrams that involve $b_1$ and $b_2$  are trivial and contribute $b_1\chi^\dagger\chi\bar u\gamma^0 u/M^2$ and $b_2\chi^\dagger\sigma^i\chi\bar u\gamma^i\gamma^5u/M^2$ to the effective theory amplitude.

\subsection{Example calculation} 
We illustrate the calculation of such QED-NRQED loop diagrams, by considering the leading power diagrams. These are the first two diagrams in figure \ref{QN_Diagrams}. We will refer to the first diagram as ``direct" and the second as ``crossed".  These diagrams arise from the NRQED field coupling in the first term of (\ref{Lagrangian2}). Applying the Feynman rules in appendix \ref{app:FR} we obtain 
\begin{equation}
{\cal M}=\alpha^2Q_\ell^2Z^2\chi^\dagger\chi\left[\bar uu\left(I_D^m+I_C^m\right)+\bar u\gamma^0u\left(I_D^0+I_C^0\right)\right],
\end{equation}
where we have defined the integrals 
\begin{equation}\label{LPintegrals}
I_{D,C}^{m,0}=(-i)(4\pi)^2\int\dfrac{d^4l}{(2\pi)^4}\dfrac{\{m,m-l^0\}}{(l^2-2ml^0+i\epsilon)(l^2-\lambda^2+i\epsilon)^2\left(\pm l^0-\dfrac{\vec l^{\,\,2}}{2M}+i\epsilon\right)}.
\end{equation}
Our notation is such that the numerator $m$ ($m-l^0$) corresponds to the superscript $m$ ($0$) and the plus (minus) sign in the last term in the denominator corresponds to the subscript $D$ ($C$).

The presence of both $l^0$ and $\vec l^{\,\,2}$ in the denominator implies that the standard covariant Feynman integral technics cannot be used. Also, unlike pure NRQED loop diagrams, see e.g. \cite{Kinoshita:1995mt}, we also have relativistic fermion propagators.  Furthermore, we need only terms that are singular in $\lambda$ and ${\cal O}(\lambda^0)$ terms up to power $1/M^2$. We will present two methods to calculate these integrals.   

In the first method we use contour integration for the $l^0$ integrals. Since we have no external 3-momenta in the problem, the angular integration is also immediate. We are left with the integral over $|\vec l|$. It is convenient at this stage to rescale $\lambda,m$, and
$|\vec l|$ by $M$ so the integral depends only on $\lambda/M$ and $m/M$. We use a simple version of the method of regions \cite{Becher:2014oda}. We divide the integration domain to a region where $|\vec l |\sim\lambda$ and a region where  $|\vec l|\gg\lambda$. The two regions are separated by an arbitrary scale that cancels when we add the two together. For the lower region  we expand in $|\vec l |\sim\lambda$. For the upper region we can set $\lambda$ to zero. We add the two regions, confirm that the arbitrary cutoff cancels, and expand the ${\cal O}(\lambda^0)$ terms up to $1/M^2$. The results of the integrals in (\ref{LPintegrals}) are 
\begin{eqnarray}\label{FirstMethod}
I_C^m&=&\dfrac1{\lambda^2}- \dfrac{\pi}{ 8 m \lambda}-\dfrac{3 \pi}{8 M \lambda}+\dfrac1{6 m^2}+\dfrac5{12 m M}+ \dfrac{\log(m/\lambda)}{2mM}+ \dfrac{5\log(m/\lambda)}{4M^2}+\dfrac5{12M^2},\nonumber\\
I_D^m&=&\dfrac{2M m \pi}{(m+M)\lambda^3}-\dfrac1{\lambda^2}+\dfrac{\pi(3m^3+5m^2-mM^2+M^3)}{8mM(m+M)^2\lambda}-\dfrac1{6 m^2}+\dfrac5{12 m M}\nonumber\\
&&+ \dfrac{\log(m/\lambda)}{2mM}- \dfrac{5\log(m/\lambda)}{4M^2}-\dfrac5{12M^2},\nonumber\\
I_C^0&=&\dfrac1{\lambda^2}- \dfrac{5\pi}{ 8 m \lambda}-\dfrac{3 \pi}{8 M \lambda}+\dfrac2{3 m^2}+\dfrac5{12 m M}+\dfrac{\log(m/\lambda)}{m^2}+ \dfrac{2\log(m/\lambda)}{mM}+ \dfrac{5\log(m/\lambda)}{4M^2}\nonumber\\&&-\dfrac{15\log(4M/m)}{8M^2}+\dfrac{161}{48M^2},\nonumber\\
I_D^0&=&\dfrac{2M m \pi}{(m+M)\lambda^3}-\dfrac1{\lambda^2}+\dfrac{\pi(3m+M)(m^2-3M^2)}{8mM(m+M)^2\lambda}-\dfrac2{3 m^2}+\dfrac5{12 m M}-\dfrac{\log(m/\lambda)}{m^2}\nonumber\\
&&+ \dfrac{2\log(m/\lambda)}{mM}- \dfrac{5\log(m/\lambda)}{4M^2}+\dfrac{15\log(4M/m)}{8M^2}-\dfrac{161}{48M^2}.
\end{eqnarray}    

In the second method we expand the NRQED propagator before we calculate the integrals:
\begin{equation}\label{expanded} 
\left(\pm l^0-\dfrac{\vec l^{\,\,2}}{2M}+i\epsilon\right)^{-1}=\pm\dfrac1{l^0}+\dfrac{\vec l^{\,\,2}}{2(l^0)^2M}\pm\dfrac{(\vec l^{\,\,2})^2}{4(l^0)^3M^2}+{\cal O}\left(\dfrac1{M^3}\right).
\end{equation}
We then use contour integration for $l^0$ integrals, perform the angular integration and integrate over $|\vec l|$. Since $\vec l^{\,\,2}$ appears only in the numerator after the expansion, the last integration is easier to do than in the previous method . Furthermore, we automatically generate the series in $1/M$. The price we pay is that the IR singular terms in $\lambda$ are expanded in $1/M$ too. This is not a problem in practice since such singular terms cancel in the matching. The presence of the $\vec l^{\,\,2}$ in the numerator might make the integral UV divergent. We regularize those by a cutoff  $\Lambda\sim M$.  Using the second method the integrals in (\ref{LPintegrals}) are 
\begin{eqnarray}
I_C^m&=&\dfrac1{\lambda^2}- \dfrac{\pi}{ 8 m \lambda}-\dfrac{3 \pi}{8 M \lambda}+\dfrac1{6 m^2}+\dfrac5{12 m M}+ \dfrac{\log(m/\lambda)}{2mM}+ \dfrac{5\log(m/\lambda)}{4M^2}+\dfrac5{12M^2},\nonumber\\
I_D^m&=&\dfrac{2m \pi}{\lambda^3}\left(1-\dfrac{m}{M}+\dfrac{m^2}{M^2}\right)-\dfrac1{\lambda^2}+\dfrac{\pi}{ 8 m \lambda}-\dfrac{3 \pi}{8 M \lambda}+\dfrac{5 m\pi}{4 M^2 \lambda}-\dfrac1{6 m^2}+\dfrac5{12 m M}+ \dfrac{\log(m/\lambda)}{2mM}\nonumber\\
&&-\dfrac{5\log(m/\lambda)}{4M^2}-\dfrac5{12M^2}.
\end{eqnarray} 
\begin{eqnarray}
I_C^0&=&\dfrac1{\lambda^2}- \dfrac{5\pi}{ 8 m \lambda}-\dfrac{3 \pi}{8 M \lambda}+\dfrac2{3 m^2}+\dfrac5{12 m M}+\dfrac{\log(m/\lambda)}{m^2}+ \dfrac{2\log(m/\lambda)}{mM}+ \dfrac{5\log(m/\lambda)}{4M^2}\nonumber\\&&-\dfrac{15\log(2\Lambda/m)}{8M^2}+\dfrac{113}{48M^2},\nonumber\\
I_D^0&=&\dfrac{2m \pi}{\lambda^3}\left(1-\dfrac{m}{M}+\dfrac{m^2}{M^2}\right)-\dfrac1{\lambda^2}+ \dfrac{5\pi}{ 8 m \lambda}-\dfrac{11 \pi}{8 M \lambda}+\dfrac{9 m\pi}{4 M^2 \lambda}-\dfrac2{3 m^2}+\dfrac5{12 m M}-\dfrac{\log(m/\lambda)}{m^2}\nonumber\\&&+ \dfrac{2\log(m/\lambda)}{mM}- \dfrac{5\log(m/\lambda)}{4M^2}+\dfrac{15\log(2\Lambda/m)}{8M^2}-\dfrac{113}{48M^2}.
\end{eqnarray}    
It is easy to check that these expressions correspond to the $1/M$ expansion of (\ref{FirstMethod}) by choosing the value of $\Lambda$ appropriately. Due to its simplicity,  we will use the second method in the results presented below.

We note that most of the direct and crossed terms that contain even inverse powers of $M$ differ in sign, while those with odd powers have the same sign. This is easily understood from (\ref{expanded}) where such a pattern appears in the expanded propagator. For singular terms such relations may not hold. In particular, the direct diagram alone has a $1/\lambda^{3}$ singularity.

Because of such relations, many terms cancel if we add the direct and crossed integrals. This simplifies the structure of the spin-independent parts of the amplitude that typically contain a sum of the direct and crossed integrals. Furthermore, since the amplitude scales as inverse mass squared, the terms proportional to $1/M$ will be multiplied by $1/m$ or $1/\lambda$. Thus they are IR divergent and cancel in the matching. Terms proportional to $1/M^2$ typically cancel in the sum and do not contribute to the matching coefficient of the spin-independent terms. A non-zero contribution will have to come from the full theory amplitude. This has important implications for what follows. The spin-dependent parts of the amplitude typically contain the difference of the direct and crossed integrals. Thus  terms proportional to $1/M^2$ would not cancel and may contribute to the matching coefficient of the spin-dependent terms.

The  leading power amplitude, corresponding to the first two diagrams in figure \ref{QN_Diagrams}, is 
\begin{eqnarray}
\dfrac{{\cal M}^{\mbox{\scriptsize EFT}}_{\mbox{\scriptsize L.P.}}}{\alpha^2Q_\ell^2Z^2}&=&
\chi^\dagger\chi\,\bar u\gamma^0u\left[\dfrac{2m \pi}{\lambda^3}\left(1-\dfrac{m}{M}+\dfrac{m^2}{M^2}\right)-\dfrac{7 \pi}{4 M \lambda}+\dfrac{9 m\pi}{4 M^2 \lambda}+\dfrac{4\log(m/\lambda)}{mM}+\dfrac{5}{6mM}\right]\nonumber\\
&+&\chi^\dagger\chi\,\bar uu\left[\dfrac{2m \pi}{\lambda^3}\left(1-\dfrac{m}{M}+\dfrac{m^2}{M^2}\right)-\dfrac{3 \pi}{4 M \lambda}+\dfrac{5 m\pi}{4 M^2 \lambda}+ \dfrac{\log(m/\lambda)}{mM}+\dfrac{5}{6mM}\right].
\end{eqnarray} 

We note that the entire leading power effective field theory contributions is given by terms that diverge when $\lambda$ and/or $m$ go to zero. Such IR terms will cancel in the matching.  

\subsection{Final result}
Using similar methods we calculate the other  diagrams in Figure \ref{QN_Diagrams}. The sum of all the diagrams is
\begin{align}\label{Total_EFT}
\dfrac{{\cal M}^{\mbox{\scriptsize EFT}}}{\alpha^2Q_\ell^2}&=\chi^\dagger\chi\bar u u\Bigg[Z^2\left(\dfrac{2m \pi}{\lambda^3}-\dfrac{2m^2 \pi}{M\lambda^3}+\dfrac{2m^3 \pi}{M^2\lambda^3}+\dfrac{3 \pi}{4M\lambda}+\dfrac{3 m\pi}{4M^2\lambda}-\dfrac{2}{3mM}-\dfrac{2\log(m/\lambda)}{mM}\right)\nonumber\\
&\qquad\qquad\quad{}-c_F^2 \dfrac{m\pi}{M^2\lambda}-c_DZ \dfrac{m\pi}{2M^2\lambda}\Bigg]+\nonumber\\
&+\chi^\dagger\chi\bar u\gamma^0 u\Bigg[Z^2\left(\dfrac{2m \pi}{\lambda^3}-\dfrac{2m^2 \pi}{M\lambda^3}+\dfrac{2m^3 \pi}{M^2\lambda^3}-\dfrac{5 \pi}{4M\lambda}+\dfrac{3 m\pi}{4M^2\lambda}-\dfrac{2}{3mM}+\dfrac{4\log(m/\lambda)}{mM}\right)\nonumber\\
&\qquad\qquad\qquad{}+c_F^2 \dfrac{m\pi}{M^2\lambda}-c_DZ \dfrac{m\pi}{2M^2\lambda}\Bigg]+\nonumber\\
&+\chi^\dagger\sigma^i\chi\bar u\left(\frac{i}2\epsilon^{ijk}\gamma^j\gamma^k\right)u\Bigg[c_FZ \left(\dfrac{2m\pi}{3M^2\lambda}-\dfrac{2\log(m/\lambda)}{M^2}\right)+c_F^2 \left(\dfrac{m\pi}{3M^2\lambda}-\dfrac{\log(m/\lambda)}{M^2}\right)\nonumber\\
&\qquad\qquad\qquad\qquad\qquad\quad\,{}+c_SZ \left(-\dfrac1{M^2}+\dfrac{\log(m/\lambda)}{M^2}\right)\Bigg]+\nonumber\\
&+\chi^\dagger\sigma^i\chi\bar u\gamma^i\gamma^5u\Bigg[c_FZ \left(-\dfrac{4\pi}{3M\lambda}+\dfrac{2m\pi}{3M^2\lambda}+\dfrac{2\log(2\Lambda/\lambda)}{M^2}+\dfrac{2\log(2\Lambda/m)}{M^2}-\dfrac{16}{3M^2}\right)\nonumber\\
&\qquad\qquad\qquad\quad{}+c_F^2 \left(-\dfrac{m\pi}{3M^2\lambda}+\dfrac{\log(m/\lambda)}{M^2}-\dfrac{\log(2\Lambda/m)}{2M^2}-\dfrac{1}{12M^2}\right)\nonumber\\
&\qquad\qquad\qquad\quad{}+c_SZ \left(-\dfrac{\log(m/\lambda)}{M^2}-\dfrac{3\log(2\Lambda/m)}{2M^2}+\dfrac{13}{12M^2}\right)\Bigg].
\end{align}  
Notice that  the sum of all spin-independent terms diverges when $\lambda$ and/or $m$ go to zero. Such IR terms will cancel in the matching.  The spin-dependent terms contain non-IR terms that can contribute to the matching coefficients, unless they match the full theory terms. The results of  the QED-NRQED calculation in the Coulomb gauge are given in appendix \ref{app:CG}. 

\section{Full Theory Calculation}\label{sec:Full}  

\subsection{Toy example: non-relativistic point particle}
For the full theory calculation we find it instructive to consider first the toy example of the point-particle ``proton". While obviously this is not a realistic model, it exhibits some of the features we will encounter later in the rigorous description of the proton in terms of the hadronic tensor.  

We consider the interaction of a relativistic point particle, e.g. a  lepton, with a non-relativistic point-particle ``proton".  There are two diagrams that contribute to the amplitude, a ``direct" diagram and a ``crossed" diagram. The amplitude is 
\begin{equation}
i{\cal M}=e^4Z^2Q^2_\ell\int \dfrac{d^4l}{(2\pi)^4}\dfrac1{(l^2-\lambda^2)^2}\dfrac1{l^2+2mv\cdot l}\left[\dfrac{A^{\mu\nu}(l)B_{\mu\nu}(-l)}{l^2-2Mv\cdot l}+\dfrac{A^{\mu\nu}(l)B_{\nu\mu}(l)}{l^2+2Mv\cdot l}\right]
\end{equation}
where 
\begin{eqnarray}\label{AB}
A^{\mu\nu}(l)&=&\bar u(k)\gamma^\mu\left[\lslash+m(1+\gamma^0)\right]\gamma^\nu u(k)\nonumber\\
B^{\mu\nu}(l)&=&\bar u(p)\gamma^\mu\left[\lslash+M(1+\gamma^0)\right]\gamma^\nu u(p),
\end{eqnarray}
and $v=(1,\vec 0)$, $k=mv$, $p=Mv$. From now on we will suppress the argument of $u(k)$.

Since we will match onto QED-NRQED, we use non-relativistic normalization for the proton spinors $u(p)^\dagger u(p)=1$. We also use the Dirac representation of the $\gamma$ matrices 
$$
\gamma^0=\begin{pmatrix} \,1&0 \\ \,0&-1\, \end{pmatrix},\quad \gamma^i=\begin{pmatrix} 0&\sigma^i \\ -\sigma^i & 0 \end{pmatrix}, 
$$
and take $u(p)\to(\chi\,\, 0)^T$. 

The components of $A^{\mu\nu}$ are 
\begin{eqnarray}
&&A^{00}(l)=\bar u\left[\gamma^0\left(l^0+m\right)+l^i\gamma^i+m\right]u\nonumber\\
&&A^{0i}(l)=\bar u\left[\gamma^i\left(l^0+m\right)+m\gamma^0\gamma^i+l^j\left(\delta^{ij}\gamma^0-i\epsilon^{ijk}\gamma^k\gamma^5\right)\right]u\nonumber\\
&&A^{i0}(l)=\bar u\left[\gamma^i\left(l^0+m\right)+m\gamma^i\gamma^0+l^j\left(\delta^{ij}\gamma^0+i\epsilon^{ijk}\gamma^k\gamma^5\right)\right]u\nonumber\\
&&A^{ij}(l)=\bar u\left[\left(\delta^{ij}\gamma^0+i\epsilon^{ijk}\gamma^k\gamma^5\right)\left(l^0+m\right)-l^k\gamma^i\gamma^k\gamma^j+m\gamma^i\gamma^j\right]u,
\end{eqnarray}
where we have used the identity $\gamma^0\gamma^i\gamma^j=-\left(\delta^{ij}\gamma^0+i\epsilon^{ijk}\gamma^k\gamma^5\right)$.
The components of $B^{\mu\nu}$ are 
\begin{eqnarray}
&&B^{00}(l)=\chi^\dagger\chi(2M+l^0),\qquad B^{0i}=\chi^\dagger\left(l^i-i\epsilon^{ijk}l^j\sigma^k\right)\chi\nonumber\\
&&B^{i0}(l)=\chi^\dagger\left(l^i+i\epsilon^{ijk}l^j\sigma^k\right)\chi, \quad B^{ij}=l^0\chi^\dagger\left(\delta^{ij}+i\epsilon^{ijk}\sigma^k\right)\chi.
\end{eqnarray}
Neglecting terms linear in $\vec l$ that integrate to zero we find
\begin{eqnarray}
&&A^{\mu\nu}(l)B_{\nu\mu}(l)=\chi^\dagger\chi\bar uu\,2m(M-l^0)+\chi^\dagger\chi\bar u\gamma^0u\left[2(m+l^0)(M+2l^0)-2\vec l^{\,\,2}\right]\nonumber\\
&+&\chi^\dagger\sigma^i\chi\bar u\left(\frac{i}2\epsilon^{ijk}\gamma^j\gamma^k\right)u(-2ml^0)+\chi^\dagger\sigma^i\chi\bar u\gamma^i\gamma^5u\left[-\dfrac43\vec l^{\,\,2}+2l^0(m+l^0)\right]\nonumber\\
&&A^{\mu\nu}(l)B_{\mu\nu}(-l)=\chi^\dagger\chi\bar uu\,2m(M+l^0)+\chi^\dagger\chi\bar u\gamma^0u\left[2(m+l^0)(M-2l^0)+2\vec l^{\,\,2}\right]\nonumber\\
&+&\chi^\dagger\sigma^i\chi\bar u\left(\frac{i}2\epsilon^{ijk}\gamma^j\gamma^k\right)u(-2ml^0)+\chi^\dagger\sigma^i\chi\bar u\gamma^i\gamma^5u\left[-\dfrac43\vec l^{\,\,2}+2l^0(m+l^0)\right],
\end{eqnarray}
where for integrals over $l^il^j$ we replace $l^il^j\to\delta^{ij}\vec l^{\,\,2}/3$. We need the following integrals. 
\begin{equation}
I(M),I^0(M),I^{00}(M), \tilde{I}(M)=(-i)(4\pi)^2\int \dfrac{d^4l}{(2\pi)^4}\dfrac{\left\{1,\,l^0,\,l^0l^0,\,\vec l^{\,\,2}\right\}}{(l^2-\lambda^2)^2(l^2+2mv\cdot l)(l^2+2Mv\cdot l)}.
\end{equation}
To calculate them we use the partial fractioning identity 
\begin{equation}
\frac1{(l^2+2mv\cdot l)(l^2+2Mv\cdot l)}=\dfrac1{2(M-m)}\dfrac1{v\cdot l}\left(\frac1{l^2+2mv\cdot l}- \frac1{l^2+2Mv\cdot l}\right)\,,
\end{equation}
and define the integrals 
\begin{equation}
i(m),i^0(m),i^{00}(m), \tilde{i}(m)=(-i)(4\pi)^2\int \dfrac{d^4l}{(2\pi)^4}\dfrac{\left\{1,\,l^0,\,l^0l^0,\,\vec l^{\,\,2}\right\}}{(l^2-\lambda^2)^2(v\cdot l)(l^2+2mv\cdot l)}. 
\end{equation}
We can now express $I,I^0,I^{00}, \tilde{I}$ in terms of  $i,i^0,i^{00}, \tilde{i}$: 
\begin{equation}
I(M)=\dfrac{i(m)-i(M)}{2(M-m)}, I^0(M)=\dfrac{i^0(m)-i^0(M)}{2(M-m)},
I^{00}(M)=\dfrac{i^{00}(m)-i^{00}(M)}{2(M-m)},\tilde{I}(M)=\dfrac{\tilde{i}(m)-\tilde{i}(M)}{2(M-m)}.
\end{equation}
In calculating the $i$ integrals it is convenient to combine denominators first via \cite{Manohar:2000dt} 
\begin{equation}
\frac1{(l^2+2mv\cdot l)(v\cdot l)}=\int_0^\infty dy\dfrac{2}{\left(l^2+2mv\cdot l+2yv\cdot l\right)^2}.
\end{equation}
Calculating  $i,i^0,i^{00}, \tilde{i}$ we find
\begin{eqnarray}
i(m)&=&\dfrac{\pi\left(m-\sqrt{m^2}\right)}{m\lambda^3}+\dfrac1{m\lambda^2}-\dfrac{\sqrt{m^2}\, \pi}{8 m^3\lambda}+\dfrac1{6m^3}\nonumber\\
i^0(m)&=&-\dfrac{\pi}{2\sqrt{m^2}\lambda}+\dfrac{1+\log m^2-\log\lambda^2}{2m^2}\nonumber\\
i^{00}(m)&=&\dfrac{-2+\log m^2-\log\lambda^2}{2m}\nonumber\\
\tilde{i}(m)&=&\dfrac{\pi\left(m-\sqrt{m^2}\right)}{m\lambda}+\dfrac{3\log m^2-3\log\lambda^2}{2m}.
\end{eqnarray}
Using the expressions above, the amplitude is 
\begin{eqnarray}\label{QN_pp_FG} 
&&\dfrac{{\cal M}^{\scriptsize\mbox{p.p.}}}{Q_l^2Z^2\alpha^2}=\chi^\dagger\chi\bar uu\left[\dfrac{2mM\pi}{(m+M)\lambda^3}+\dfrac{3\pi}{4(m+M)\lambda}+\dfrac2{mM}\left(\log \lambda-\dfrac13-\dfrac{m^2\log M-M^2\log m}{m^2-M^2}\right)\right]\nonumber\\
&+&\chi^\dagger\chi\bar u\gamma^0u\left[\dfrac{2mM\pi}{(m+M)\lambda^3}-\dfrac{5\pi}{4(m+M)\lambda}+\dfrac2{mM}\left(-2\log \lambda-\dfrac13+\dfrac{2(m^2\log M-M^2\log m)}{m^2-M^2}\right)\right]\nonumber\\
&+&\chi^\dagger\sigma^i\chi\bar u\left(\frac{i}2\epsilon^{ijk}\gamma^j\gamma^k\right)u\left[\dfrac{m\pi}{M(m+M)\lambda}+\dfrac{2\log\lambda-1}{M^2}+\dfrac{2\left(M^2\log m-m^2\log M\right)}{M^2(m^2-M^2)}\right]\nonumber\\
&+&\chi^\dagger\sigma^i\chi\bar u\gamma^i\gamma^5u
\left[-\dfrac{\pi}{M\lambda}-\dfrac{\pi}{3(m+M)\lambda}+\dfrac{1+2\log M-2\log\lambda}{M^2}\right].
\end{eqnarray}
We will see in appendix \ref{app:CG} that in Coulomb gauge this amplitude (\ref{QN_pp_FG}) is different.  In the next section we will use this result to extract $b_1$ and $b_2$ for the non-relativistic point particle case.

As a check, we can take the non-relativistic limit for the lepton, i.e.  $\bar uu,\bar u\gamma^0u\to \chi_\ell^\dagger\chi_\ell$, and $\bar u\left(\frac{i}2\epsilon^{ijk}\gamma^j\gamma^k\right)u, \bar u\gamma^i\gamma^5u\to \chi_\ell^\dagger\sigma^i\chi_\ell$. The resulting  amplitude is 
\begin{eqnarray}\label{NN_pp_FG}
\dfrac{{\cal M}^{\scriptsize\mbox{p.p.}}}{Q_l^2Z^2\alpha^2}&=&\chi_p^\dagger\chi_p\chi_\ell^\dagger\chi_\ell\left[\dfrac{4mM\pi}{(m+M)\lambda^3}-\dfrac{\pi}{2(m+M)\lambda}+\dfrac2{mM}\left(-\log \lambda-\dfrac23+\dfrac{m^2\log M-M^2\log m}{m^2-M^2}\right)\right]\nonumber\\
&+&\chi_p^\dagger\sigma^i\chi_p\chi_\ell^\dagger\sigma^i\chi_\ell\left[-\dfrac{4\pi}{3(m+M)\lambda}+\dfrac{2\log (m/M)}{m^2-M^2}\right],
\end{eqnarray}
which is a known result \cite{HP}. We will see in appendix \ref{app:CG} that unlike (\ref{QN_pp_FG}), we get the same result for (\ref{NN_pp_FG}) in Coulomb gauge. 
 
\subsection{The proton}\label{subsec:proton}
For the proton the forward lepton-proton ${\cal O}(Z^2\alpha^2)$ amplitude can be expressed in terms of the hadronic tensor 
\begin{equation}\label{Wdefined}
W^{\mu\nu}(p,q)=i\int d^4x\, e^{iqx}\langle p,s|T\left\{J^\mu_{\mbox{\scriptsize e.m.}}(x)J_{\mbox{\scriptsize e.m.}}^\nu(0)\right\}|p,s \rangle\,,
\end{equation}
that encodes the two-photon interaction of the proton. Here $p$ is the proton momentum and $s$ is its spin. As we review in appendix \ref{app:HT}, $W^{\mu\nu}(p,q)$ can be expressed as
\begin{eqnarray}\label{Tensordecomposition}
W^{\mu\nu}(p,q)=\dfrac1{2\mproton}\bar u_p(p,s) &&\Bigg[\left( - g^{\mu\nu} + \dfrac{q^\mu q^\nu}{ q^2} \right) W_1+
\left( p^\mu - \dfrac{p\cdot q \,q^\mu }{q^2} \right) 
\left( p^\nu - \dfrac{p\cdot q \, q^\nu} {q^2} \right) W_2\nonumber\\
&&+\bigg( [\gamma^\nu,\qslash]\,p^\mu- [\gamma^\mu,\qslash]\,p^\nu+[\gamma^\mu,\gamma^\nu]\,p\cdot q \bigg) H_1\nonumber\\
&&+\bigg( [\gamma^\nu,\qslash]\,q^\mu- [\gamma^\mu,\qslash]\,q^\nu+[\gamma^\mu,\gamma^\nu]\,q^2  \bigg) H_2\Bigg]u_p(p,s).
\end{eqnarray} 
The four scalar functions $W_{1},W_{2},H_{1},H_{2}$ depend on the variables $\nu=2p\cdot q$ and $Q^2=-q^2$. Our definition of $H_{1,2}$ is related to $G_{1,2}$ of \cite{Bjorken:1966jh}, via $H_i=-MG_i$. Translation invariance implies that $W^{\mu\nu}(p,q)=W^{\nu\mu}(p,-q)$. As a result, $W_1$, $W_2$, $H_1$  are 
even functions of $\nu$, and $H_2$ is odd; see appendix \ref{app:HT}. For a point particle, $W_1=2\nu^2/(Q^4-\nu^2)$, $W_2=8Q^2/(Q^4-\nu^2)$, $H_1=-2Q^2/(Q^4-\nu^2)$, and $H_2=0$.

In terms of the hadronic tensor,  the amplitude in Feynman gauge is 
\begin{equation}\label{full}
i{\cal M}_{\mbox{\scriptsize Full}}=-Q_\ell^2e^4\int\dfrac{d^4l}{(2\pi)^4}\dfrac{\bar u\gamma_\mu(\kslash-\lslash+m)\gamma_\nu u}{(k-l)^2-m^2}\left(\dfrac{1}{l^2-\lambda^2}\right)^2W^{\mu\nu}(p,l).
\end{equation}

We insert (\ref{Tensordecomposition}) into (\ref{full}) and set the lepton 3-momentum to zero, i.e. $k=(m,\vec 0)$ and work in the rest frame of proton, i.e. $p=(M,\vec 0)$. To match onto NRQED, the proton Dirac spinors in (\ref{Tensordecomposition}) should be used with non-relativistic normalization $u(p)^\dagger u(p)=1$ and we replace $u_p\to(\chi\,\,0)^T$, where $\chi$ is a two-component spinor. After these simplifications we find
\begin{eqnarray}\label{master_formula}
i{\cal M}&=&-Q_\ell^2e^4\int\dfrac{d^4l}{(2\pi)^4}\dfrac{1}{l^2-2ml^0}\left(\dfrac{1}{l^2-\lambda^2}\right)^2\dfrac1{2M}\times\nonumber\\
&&\Bigg\{-W_1\,\chi^\dagger\chi\left[\bar uu\cdot 3m+\bar u\gamma^0u\cdot (2l^0-m)\right]+W_2\,\chi^\dagger\chi\left(\bar uu+\bar u\gamma^0u\right)M^2m\left[1-\dfrac{\left(l^0\right)^2}{l^2}\right]\nonumber\\
&&+H_1\,\chi^\dagger\sigma^i\chi\left[\bar u\gamma^i\gamma^5u\left(\dfrac83M{\vec l}^{\,\,2}+4M(m-l^0)l^0\right)+\bar u\left(\frac{i}{2}\epsilon^{ijk}\gamma^j\gamma^k\right)u\left(-4Mml^0\right)\right]\nonumber\\
&&+H_2\,\chi^\dagger\sigma^i\chi
\left[\bar u\gamma^i\gamma^5u\left(\dfrac83m{\vec l}^{\,\,2}+4(m-l^0)l^2\right)
-\bar u\left(\frac{i}{2}\epsilon^{ijk}\gamma^j\gamma^k\right)u\left(\dfrac83 m{\vec l}^{\,\,2}+4ml^2\right)
\right]\Bigg\}.\nonumber\\
\end{eqnarray}
In deriving this equation, we used that $W_1$ and $W_2$ are even functions of $\nu=2Ml^0$. As a check, we  insert the point particle values of $W_1,W_2,H_1,H_2$ into (\ref{master_formula}) and reproduce the result of (\ref{QN_pp_FG}).  

To proceed, we use dispersion relations for the scalar functions $W_{1},W_{2},H_{1},H_{2}$.  For a given value of $Q^2$ the singularities of these functions lie on the real $\nu$ axis and are symmetric under $\nu\to-\nu$ see, e.g., \cite{Manohar:2000dt}. These consist of a pole at $\nu=\pm Q^2$, corresponding to proton state, and a cut that starts at $\nu_{\mbox{\scriptsize cut}}(Q^2)=Q^2+2m_\pi \mproton+m_\pi^2$, the threshold for a production of a proton and pion pair\footnote{One might be concerned that in the matching one could have terms of the form $m_\mu/m_\pi$, that are not suppressed, instead of $m_\mu/M$. We think this is unlikely.  Since the proton and pion appear together in the continuum contribution, expanding in inverse powers of the proton mass  will only generate positive powers of the pion mass. The subtraction term, $W_1(0,Q^2)$, that does not formally separate to proton and continuum states is discussed at length in \cite{Hill:2016bjv}.}. Using the symmetry of the scalar functions under $\nu\to-\nu$ we can express them as integrals over the positive value of $\nu$ starting at $\nu_0\equiv Q^2$.

$W_{1}$ satisfies a subtracted dispersion relation in $\nu$, 
\begin{equation}\label{W1disp}
W_1(\nu,Q^2)-W_1(0,Q^2)=\dfrac{2\nu^2}{\pi}\int_{\nu_0}^\infty d\nu^\prime\dfrac{\mbox{Im } W_1(\nu^\prime,Q^2)}{\nu^\prime\left(\nu^{\prime2}-\nu^2\right)},
\end{equation}
see \cite{Hill:2016bjv} for a proof that utilizes the optical and low energy theorems. We assume that $W_{2},H_{1},H_{2}$ satisfy unsubtracted dispersion relations
\begin{eqnarray}\label{W2disp}
W_2(\nu,Q^2)&=&\dfrac{2}{\pi}\int_{\nu_0}^\infty d\nu^\prime\,\nu^\prime\dfrac{\mbox{Im } W_2(\nu^\prime,Q^2)}{\nu^{\prime2}-\nu^2}\nonumber\\
H_1(\nu,Q^2)&=&\dfrac{2}{\pi}\int_{\nu_0}^\infty d\nu^\prime\,\nu^\prime\dfrac{\mbox{Im } H_1(\nu^\prime,Q^2)}{\nu^{\prime2}-\nu^2}\nonumber\\
H_2(\nu,Q^2)&=&\dfrac{2\nu}{\pi}\int_{\nu_0}^\infty d\nu^\prime\,\dfrac{\mbox{Im } H_2(\nu^\prime,Q^2)}{\nu^{\prime2}-\nu^2}.
\end{eqnarray}

From the definition of $W^{\mu\nu}$ (\ref{Wdefined}) one can obtain the identity \cite{Manohar:2000dt}
\begin{eqnarray}\label{ImW}
2\,\mbox{Im}\, W^{\mu\nu}&=&\sum_X \langle p,s|J^\mu|X\rangle\langle X |J^\nu|p,s\rangle (2\pi)^4 \delta^{(4)}\left(q-p_X+p\right)\nonumber\\
&+&\sum_X \langle p,s|J^\nu|X\rangle\langle X |J^\mu|p,s\rangle (2\pi)^4 \delta^{(4)}\left(q+p_X-p\right),
\end{eqnarray}
where $p_X$ is the four-momentum of the state $X$. The summation includes phase space integrals and sum over spins of the state $X$. This equation allows us to separate the proton and the continuum contribution in the imaginary part of $W^{\mu\nu}$. The proton contribution is given in terms of form factors, and the continuum contribution in terms of  inelastic structure functions.

Using the dispersion relations we can express the scalar functions as \cite{Hill:2011wy}
\begin{eqnarray}\label{decomposition}
W_1(\nu,Q^2)&=&W_1(0,Q^2)+W_1^{p,1}(\nu,Q^2)+W_1^{c,1}(\nu,Q^2)\nonumber \\
W_2(\nu,Q^2)&=&W_2^{p,0}(\nu,Q^2)+W_2^{c,0}(\nu,Q^2)\nonumber \\
H_1(\nu,Q^2)&=&H_1^{p,0}(\nu,Q^2)+H_1^{c,0}(\nu,Q^2)\nonumber \\
H_2(\nu,Q^2)&=&H_2^{p,0}(\nu,Q^2)+H_2^{c,0}(\nu,Q^2),
\end{eqnarray}
where the superscript numbers denote the number of subtractions, the superscript $p$ is the proton contribution, and the superscript $c$ is the continuum contribution. 

Using (\ref{FF}) we can express the proton contribution in (\ref{ImW}) in terms of the form factors
\begin{eqnarray}
2\,\mbox{Im}\,W^{\mu\nu}_{\mbox{\scriptsize proton}}&=& 2\pi\delta(\nu-Q^2)\,\bar u\left(F_1\gamma^\mu-\dfrac{i\sigma^{\mu\alpha}q_\alpha}{2M}F_2\right)\left(\pslash+\qslash+M\right)\left(F_1\gamma^\nu+\dfrac{i\sigma^{\nu\beta}q_\beta}{2M}\right)u\nonumber\\
&+&2\pi\delta(\nu+Q^2)\,\bar u\left(F_1\gamma^\nu+\dfrac{i\sigma^{\nu\alpha}q_\alpha}{2M}F_2\right)\left(\pslash+\qslash+M\right)\left(F_1\gamma^\mu-\dfrac{i\sigma^{\mu\beta}q_\beta}{2M}\right)u,
\end{eqnarray}
where $F_i\equiv F_i(q^2)=F_i(-Q^2)$. After simplifying the Dirac structure, as explained in appendix \ref{app:HT}, we find 
\begin{eqnarray}\label{ImWproton} 
\mbox{Im } W_1^p(\nu,Q^2)&=&\left[\pi\delta(\nu-Q^2)+\pi\delta(\nu+Q^2)\right] \left(F_1+F_2\right)^2Q^2\nonumber\\
\mbox{Im } W_2^p(\nu,Q^2)&=&\left[\pi\delta(\nu-Q^2)+\pi\delta(\nu+Q^2)\right] \left(4F_1^2+F_2^2Q^2/\mproton^2\right)\nonumber\\
\mbox{Im } H_1^p(\nu,Q^2)&=&\left[\pi\delta(\nu-Q^2)+\pi\delta(\nu+Q^2)\right] \left[-F_1(F_1+F_2)\right]\nonumber\\
\mbox{Im } H_2^p(\nu,Q^2)&=&\left[\pi\delta(\nu-Q^2)-\pi\delta(\nu+Q^2)\right] \left[F_2(F_1+F_2)/2\right].
\end{eqnarray} 
Inserting these expressions into (\ref{W1disp}) and (\ref{W2disp}) gives 
\begin{eqnarray}\label{Wproton} 
W_1^{p,1}(\nu,Q^2)&=&\dfrac{2\nu^2}{Q^4-\nu^2+i\epsilon}\left(F_1+F_2\right)^2\nonumber \\
W_2^{p,0}(\nu,Q^2)&=&\dfrac{2Q^2}{Q^4-\nu^2+i\epsilon}\left(4F_1^2+F_2^2Q^2/\mproton^2\right)\nonumber \\
H_1^{p,0}(\nu,Q^2)&=&\dfrac{2Q^2}{Q^4-\nu^2+i\epsilon}\left[-F_1(F_1+F_2)\right]\nonumber \\
H_2^{p,0}(\nu,Q^2)&=&\dfrac{\nu}{Q^4-\nu^2+i\epsilon}\left[F_2(F_1+F_2)\right].
\end{eqnarray}
$W_1^{p,1}(\nu,Q^2)$ and $W_2^{p,0}(\nu,Q^2)$ agree with \cite{Hill:2011wy}. $H_1^{p,0}(\nu,Q^2)$ and $H_2^{p,0}(\nu,Q^2)$ agree, up to an overall normalization with the ones given by Drell and Sullivan\footnote{We note that there is a typo in equations (4.2) and (4.3) of \cite{Drell:1966kk}. For the tensors to be anti-symmetric under $\mu\to\nu$ the sign in front of the last commutator should be changed from minus to plus.} in  \cite{Drell:1966kk}.

The subtraction function $W_1(0,Q^2)$ is not known exactly. In the large $Q^2$ limit it can be calculated using   the operator product expansion. The spin-0 contribution was calculated in \cite{Collins:1978hi} and corrected in \cite{Hill:2016bjv}. The spin-2 contribution was calculated in \cite{Hill:2016bjv}. For the matching we need the small $Q^2$ expression that can be calculated using NRQED \cite{Hill:2011wy} 
\begin{equation}
W_1(0,Q^2) = 2 a_p(2 + a_p) +Q^2 \left\{\dfrac{2m_p \bar\beta}{ \alpha}-\dfrac{a_p}{m_p^2}-(2/3)\left[ (1+a_p)^2 (r_M^p)^2 - (r_E^p)^2 \right] \right\} + {\cal O}(Q^4)\,,
\end{equation}
where $m_p$ is the proton mass, $F_2(0)=a_p$, $r_E^p$ and $r_M^p$ are the proton electric and magnetic charge radii, and $\bar\beta$ is the magnetic polarizability. This equation assumes $Z=1$. For the matching we will only need the first term $W_1(0,Q^2) = 2 F_2(0)\left[2F_1(0) + F_2(0)\right]+ {\cal O}(Q^2)$, where we have reinserted $F_1(0)=Z$.

To match the QED-NRQED singularities we will need a contribution to $H_1(\nu,Q^2)$ in the $Q^2\to0$ limit. From the low energy theorems of Low \cite{Low:1954kd} and Gell-Mann and Goldberger \cite{GellMann:1954kc} we know that  $H_1(\nu,Q^2)$ includes also a term $F_2(0)^2/(2M^2)$. We will follow Drell and Sullivan \cite{Drell:1966kk} and include it in the continuum contribution to $H_1(\nu,Q^2)$, namely $H_1^{c,0}(\nu,Q^2)$ of (\ref{decomposition}).

In the literature one often finds a different approach to calculate the proton contribution to $W^{\mu\nu}$. In this approach the on-shell vertex of (\ref{FF}) is inserted into the off-shell amplitude of (\ref{Wdefined}). The hadronic tensor is then calculated with Feynman rules that depend on $F_1$ and $F_2$. This approach is often called ``Born terms" or ``Born approximation", see e.g. \cite{Drell:1966kk}, although this is not the first term in a perturbative series which is the usual meaning of the Born approximation in quantum mechanics. In this approach one finds 
\begin{eqnarray}\label{Born}
W_1^{\mbox{\scriptsize``Born"}}(\nu,Q^2)&=&\dfrac{2\nu^2}{Q^4-\nu^2+i\epsilon}\left(F_1+F_2\right)^2+2F_2(2F_1+F_2)\nonumber \\
W_2^{\mbox{\scriptsize``Born"}}(\nu,Q^2)&=&\dfrac{2Q^2}{Q^4-\nu^2+i\epsilon}\left(4F_1^2+F_2^2Q^2/\mproton^2\right)\nonumber \\
H_1^{\mbox{\scriptsize``Born"}}(\nu,Q^2)&=&\dfrac{2Q^2}{Q^4-\nu^2+i\epsilon}\left[-F_1(F_1+F_2)\right]+F_2^2/2M^2\nonumber \\
H_2^{\mbox{\scriptsize``Born"}}(\nu,Q^2)&=&\dfrac{\nu}{Q^4-\nu^2+i\epsilon}\left[F_2(F_1+F_2)\right].
\end{eqnarray}
Compared to the expressions above, $W_2$ and $H_2$ are unchanged. The extra term for $W_1$ corresponds to assuming $W_1(0,Q^2)=2F_2(2F_1+F_2)$. The extra term for $H_1$ is $F_2^2/2M^2$. Interestingly its value at $Q^2=0$ corresponds to the $F_2(0)^2/(2M^2)$ term that we include in the continuum contribution.  This was also noted in \cite{Drell:1966kk}. For  matching the QED-NRQED IR singularities, the difference between the extra terms in (\ref{Born}) compared to the terms we have included above is irrelevant.  

In summary, to match  QED-NRQED IR singularities we need: (\ref{Wproton}), $W_1(0,Q^2)$ to zeroth order in $Q^2$, namely $2 F_2(0)\left[2F_1(0) + F_2(0)\right]$, and a part of $H_1^{c,0}(\nu,Q^2)$: $F_2(0)^2/(2M^2)$. In total 
\begin{eqnarray}\label{Wsing} 
W_1(\nu,Q^2)&=&\dfrac{2\nu^2}{Q^4-\nu^2+i\epsilon}\left(F_1+F_2\right)^2+2 F_2(0)\left[2F_1(0) + F_2(0)\right]+\nonumber\\
&+&W_1(0,Q^2)-2 F_2(0)\left[2F_1(0) + F_2(0)\right]+W_1^{c,1}(\nu,Q^2)\nonumber \\
W_2(\nu,Q^2)&=&\dfrac{2Q^2}{Q^4-\nu^2+i\epsilon}\left(4F_1^2+F_2^2Q^2/\mproton^2\right)+W_2^{c,0}(\nu,Q^2)\nonumber \\
H_1(\nu,Q^2)&=&\dfrac{2Q^2}{Q^4-\nu^2+i\epsilon}\left[-F_1(F_1+F_2)\right]+F_2(0)^2/2M^2+\nonumber \\
&+&H_1^{c,0}(\nu,Q^2)-F_2(0)^2/2M^2\nonumber \\
H_2(\nu,Q^2)&=&\dfrac{\nu}{Q^4-\nu^2+i\epsilon}\left[F_2(F_1+F_2)\right]+H_2^{c,0}(\nu,Q^2).
\end{eqnarray}
The second line for $W_1$ and $H_1$ in (\ref{Wsing}) does not contribute to the IR singular terms.

Using this information we return to the calculation of  (\ref{master_formula}). First, to simplify the notation we rename $l\to q$. The poles in (\ref{master_formula}) arise from the lepton propagator, the photon propagators (regularized by $\lambda$), and the poles from (\ref{Wproton}). The location of the poles is such that we can perform a Wick rotation $l^0\equiv q^0=iq_E^0$. Thus $q^2=\left(q^0\right)^2-\vec q^{\,\,2}=-(q_E^2+\vec q^{\,\,2})=-Q^2$. For spherical coordinates in four-dimensional Euclidean space one has $q_E^0=Q\cos\chi$, $\vec q=Q\sin\chi\left(\sin\theta\cos\phi,\sin\theta\sin\phi,\cos\phi\right)$, where $0\leq\chi,\theta\leq\pi$ and $0\leq\phi\leq 2\pi$. Thus $d^4l=d^4q=i\,4\pi\int_0^{\infty}dQ\,Q^3\int_{-1}^{1}dx\sqrt{1-x^2}$, where $x\equiv \cos\chi$. In terms of the new variables we have 
\begin{eqnarray}\label{FullQ}
&&i{\cal M}=-iQ_\ell^2e^4 \dfrac{4\pi}{(2\pi)^4}\int_0^\infty dQ\,Q^3\int_{-1}^{1}dx\sqrt{1-x^2}\left(\dfrac{1}{Q^2+\lambda^2}\right)^2\dfrac{(-Q^2+2imQx)}{Q^4+4m^2Q^2x^2}\,\dfrac1{2M}
\times\nonumber\\
&&\Bigg\{-W_1\,\chi^\dagger\chi\left[\bar uu\cdot 3m+\bar u\gamma^0u\cdot (2iQx-m)\right]+W_2\,\chi^\dagger\chi\left(\bar uu+\bar u\gamma^0u\right)M^2m\left(1-x^2\right)\nonumber\\
&&+H_1\,\chi^\dagger\sigma^i\chi\left[\bar u\gamma^i\gamma^5u\left(\dfrac83MQ^2(1-x^2)+4M(m-iQx)iQx\right)+\bar u\left(\frac{i}{2}\epsilon^{ijk}\bar u\gamma^j\gamma^k\right)u\left(-4MmiQx\right)\right]\nonumber\\
&&+H_2\,\chi^\dagger\sigma^i\chi
\left[\left(\bar u\gamma^i\gamma^5u-\bar u(\frac{i}{2}\epsilon^{ijk}\gamma^j\gamma^k)u\right)\left(-\dfrac43mQ^2(1+2x^2)\right)+4ixQ^3\bar u\gamma^i\gamma^5u
\right]\Bigg\}.
\end{eqnarray}
Since $\nu=2MiQx$, $W_1,W_2,H_1$ ($H_2$) are even (odd) function(s) of $x$. This implies,    
\begin{eqnarray}\label{FullSimplified}
{\cal M}&=&\dfrac{2Q_\ell^2\alpha^2}{M\pi}\int_0^\infty dQ\,Q^3\int_{-1}^{1}dx\sqrt{1-x^2}\left(\dfrac{1}{Q^2+\lambda^2}\right)^2\dfrac{1}{Q^2+4m^2x^2}\times\nonumber\\
&&\Bigg\{\chi^\dagger\chi\bar uu\Big[-3mW_1+mM^2(1-x^2)W_2\Big]+\chi^\dagger\chi\bar u\gamma^0u\Big[m(1-4x^2)W_1+mM^2(1-x^2)W_2\Big]\nonumber\\
&&+\chi^\dagger\sigma^i\chi\bar u\gamma^i\gamma^5u\left[\left(\dfrac{4MQ^2}{3}(2+x^2)+8Mm^2x^2\right )H_1+\left(-4ixQ^3-\dfrac83iQxm^2(1+2x^2)\right )H_2\right]\nonumber\\
&&+\chi^\dagger\sigma^i\chi\bar u\left(\frac{i}{2}\epsilon^{ijk}\gamma^j\gamma^k\right)u\left[-8Mm^2x^2H_1-\dfrac83iQxm^2(1+2x^2)H_2\right]\Bigg\}.
\end{eqnarray}
Notice the factor of $m$  in front of all the spin-independent terms.  We now insert the expressions from (\ref{Wsing}). The IR singular terms arise from the Taylor expansion of the form factors in $Q^2$. Expanding, integrating over $x$ and $Q$,  and expanding the result in inverse powers of $M$ gives
  
\begin{align}\label{Full_expanded}
&\dfrac{{\cal M}^{\mbox{\scriptsize Full}}_{\mbox{\scriptsize Expanded}}}{\alpha^2Q_\ell^2}=\chi^\dagger\chi\bar u u\Bigg[F_1(0)^2\left(\dfrac{2m \pi}{\lambda^3}-\dfrac{2m^2 \pi}{M\lambda^3}+\dfrac{2m^3 \pi}{M^2\lambda^3}+\dfrac{3 \pi}{4M\lambda}-\dfrac{3 m\pi}{4M^2\lambda}-\dfrac{2}{3mM}-\dfrac{2\log(m/\lambda)}{mM}\right)\nonumber\\
&\hspace{9.5em}-F_1(0)F_2(0) \dfrac{3m\pi}{M^2\lambda}-F_2(0)^2 \dfrac{m\pi}{M^2\lambda}-F_1(0)M^2 F_1^\prime(0)\dfrac{4m\pi}{M^2\lambda}\Bigg]\nonumber\\
&+\bar u \gamma^0u\chi^\dagger\chi\Bigg[F_1(0)^2\left(\dfrac{2m \pi}{\lambda^3}-\dfrac{2m^2 \pi}{M\lambda^3}+\dfrac{2m^3 \pi}{M^2\lambda^3}-\dfrac{5 \pi}{4M\lambda}+\dfrac{5 m\pi}{4M^2\lambda}-\dfrac{2}{3mM}+\dfrac{4\log(m/\lambda)}{mM}\right)\nonumber\\
&\hspace{5.5em}+F_1(0)F_2(0) \dfrac{m\pi}{M^2\lambda}+F_2(0)^2 \dfrac{m\pi}{M^2\lambda}-F_1(0)M^2 F_1^\prime(0)\dfrac{4m\pi}{M^2\lambda}\Bigg]\nonumber\\
&+\chi^\dagger\sigma^i\chi\bar u\left(\frac{i}2\epsilon^{ijk}\gamma^j\gamma^k\right)u\Bigg[F_1(0)^2\left(\dfrac{m\pi}{M^2\lambda}-\dfrac{2\log(m/\lambda)}{M^2}-\dfrac{1}{M^2}\right)+\nonumber\\
&\hspace{9.5em}+F_1(0)F_2(0) \left(\dfrac{4m\pi}{3M^2\lambda}-\dfrac{2\log(m/\lambda)}{M^2}-\dfrac{2}{M^2}\right)
&\hspace{-13em}
+F_2(0)^2 \left(\dfrac{m\pi}{3M^2\lambda}-\dfrac{\log(m/\lambda)}{M^2}\right)\Bigg]\nonumber\\
&+\chi^\dagger\sigma^i\chi\bar u\gamma^i\gamma^5u\Bigg[F_1(0)^2\left(-\dfrac{4\pi}{3M\lambda}+\dfrac{m\pi}{3M^2\lambda}+\dfrac{2\log(M/\lambda)}{M^2}+\dfrac{1}{M^2}\right)
+\nonumber\\
&\hspace{7.5em}
+F_1(0)F_2(0) \left(-\dfrac{4\pi}{3M\lambda}+\dfrac{2\log(m/\lambda)}{M^2}+\dfrac{2}{M^2}\right)+\nonumber\\
&\hspace{0em}+F_2(0)^2 \left(-\dfrac{m\pi}{3M^2\lambda}+\dfrac{\log(m/\lambda)}{M^2}-\dfrac{\log(M/m)}{2M^2}+\dfrac{3\log(Q/M)}{2M^2}+\dfrac{5}{8M^2}\right)\Bigg]+{\cal O}\left(\frac1{M^3}\right).
\end{align}
The $\log(Q/M)$ in the last line of (\ref{Full_expanded}) is a UV divergence that arises from the Taylor expansion of the form factor. Such a term would be regulated when using the full functional form of $F_2$. 

As a check, we can add the $\chi^\dagger\chi\bar u u$ and $\chi^\dagger\chi\bar u\gamma^0 u$ terms and compare the sum to the NRQED result on the left hand side of equation (7) of \cite{Hill:2011wy}. Setting $F_1(0)=1$ and expanding that result to order $1/M^2$, we find a complete agreement.  As another check, the terms in (\ref{Full_expanded}) proportional to $F_1(0)^2$ match the IR singular terms of (\ref{QN_pp_FG}) expanded to order $1/M^2$.

The IR divergences in (\ref{Full_expanded}) match exactly the IR divergence of (\ref{Total_EFT}), as they should. Furthermore, the $\chi^\dagger\chi\bar u u$ and $\chi^\dagger\chi\bar u\gamma^0 u$ terms include only IR divergent terms. The  $\chi^\dagger\sigma^i\chi\bar u\left(\frac{i}2\epsilon^{ijk}\gamma^j\gamma^k\right)u$ term is equal to the corresponding term in (\ref{Total_EFT}) including the non-IR term. This would lead to a zero matching coefficient at order $1/M^2$ as expected from chiral symmetry of the leptons.  In the next section we will use (\ref{Full_expanded}) to extract the matching coefficients. 

\section{Extraction of the matching coefficients at power $\bm{1/M^2}$} \label{sec:Extraction}
With the effective field theory and full theory calculations at hand, we can now extract the matching coefficients $b_1$ and $b_2$ of (\ref{contactQN}) which is the goal of this paper. We will calculate them for two cases. First, for the toy example of a non-relativistic point particle. In this case since the full theory calculation is known explicitly, we can find explicit expressions for $b_1$ and $b_2$. Second, for the case of the proton. In this case the full theory is expressed in terms of the components of the hadronic tensor and we will give an implicit expression for the matching coefficients in terms of an integral over these components.  
\subsection{Toy example: Extraction of matching coefficients for a non-relativistic point particle}
For this case the full theory amplitude is given by (\ref{QN_pp_FG}). To perform the matching we need to expand this equation to power $1/M^2$. We find  
\begin{eqnarray}\label{QN_pp_FG_expanded} 
\dfrac{{\cal M}^{\scriptsize\mbox{p.p.}}_{\mbox{\scriptsize Expanded}}}{Q_l^2Z^2\alpha^2}&=&\chi^\dagger\chi\bar uu\left[\dfrac{2m\pi}{\lambda^3}-\dfrac{2m^2\pi}{M\lambda^3}+\dfrac{2m^3\pi}{M^2\lambda^3}+\dfrac{3\pi}{4M\lambda}-\dfrac{3m\pi}{4M^2\lambda}-\dfrac2{3mM}-\dfrac{2\log(m/\lambda)}{mM}\right]\nonumber\\
&+&\chi^\dagger\chi\bar u\gamma^0u\left[\dfrac{2m\pi}{\lambda^3}-\dfrac{2m^2\pi}{M\lambda^3}+\dfrac{2m^3\pi}{M^2\lambda^3}-\dfrac{5\pi}{4M\lambda}+\dfrac{5m\pi}{4M^2\lambda}-\dfrac2{3mM}+\dfrac{4\log(m/\lambda)}{mM}\right]\nonumber\\
&+&\chi^\dagger\sigma^i\chi\bar u\left(\frac{i}2\epsilon^{ijk}\gamma^j\gamma^k\right)u\left[\dfrac{m\pi}{M^2\lambda}-\dfrac{2\log(m/\lambda)}{M^2}-\dfrac1{M^2}\right]\nonumber\\
&+&\chi^\dagger\sigma^i\chi\bar u\gamma^i\gamma^5u
\left[-\dfrac{4\pi}{3M\lambda}+\dfrac{m\pi}{3M^2\lambda}+\dfrac{2\log (M/\lambda)}{M^2}+\dfrac{1}{M^2}\right].
\end{eqnarray}
To match onto the effective theory, we need to take (\ref{Total_EFT}) and set the values of the Wilson coefficients to that of a point particle, namely,  $c^{\mbox{\scriptsize p.p.}}_F=c^{\scriptsize\mbox{p.p.}}_D=c^{\scriptsize\mbox{p.p.}}_S=F_1(0)=Z$. 

\begin{align}\label{Total_EFT_pp}
\dfrac{{\cal M}^{\mbox{\scriptsize EFT}}_{\mbox{\scriptsize p.p. limit}}}{Q_l^2Z^2\alpha^2}&=\bar u u\chi^\dagger\chi\Bigg[\dfrac{2m \pi}{\lambda^3}-\dfrac{2m^2 \pi}{M\lambda^3}+\dfrac{2m^3 \pi}{M^2\lambda^3}+\dfrac{3 \pi}{4M\lambda}-\dfrac{3 m\pi}{4M^2\lambda}-\dfrac{2}{3mM}-\dfrac{2\log(m/\lambda)}{mM}\Bigg]+\nonumber\\
&+\bar u\gamma^0 u\chi^\dagger\chi\Bigg[\dfrac{2m \pi}{\lambda^3}-\dfrac{2m^2 \pi}{M\lambda^3}+\dfrac{2m^3 \pi}{M^2\lambda^3}-\dfrac{5 \pi}{4M\lambda}+\dfrac{5 m\pi}{4M^2\lambda}-\dfrac{2}{3mM}+\dfrac{4\log(m/\lambda)}{mM}\Bigg]+\nonumber\\
&+\chi^\dagger\sigma^i\chi\bar u\left(\frac{i}2\epsilon^{ijk}\gamma^j\gamma^k\right)u\Bigg[\dfrac{m\pi}{M^2\lambda}-\dfrac{2\log(m/\lambda)}{M^2}-\dfrac1{M^2}\Bigg]+\nonumber\\
&+\chi^\dagger\sigma^i\chi\bar u\gamma^i\gamma^5u\Bigg[-\dfrac{4\pi}{3M\lambda}+\dfrac{m\pi}{3M^2\lambda}+\dfrac{2\log(2\Lambda/\lambda)}{M^2}-\dfrac{13}{3M^2}\Bigg].
\end{align}
The difference between the full and effective theory is 
\begin{equation}
{\cal M}^{\scriptsize\mbox{p.p.}}_{\mbox{\scriptsize Expanded}}-{\cal M}^{\mbox{\scriptsize EFT}}_{\mbox{\scriptsize p.p. limit}}=Q_l^2Z^2\alpha^2\dfrac1{M^2}\left[\dfrac{16}{3}+2\log\left(\dfrac{M}{2\Lambda}\right)\right]\chi^\dagger\sigma^i\chi\bar u\gamma^i\gamma^5u.
\end{equation}

As expected from chiral symmetry of the leptons, the coefficient of the structure $\chi^\dagger\chi\bar u u$ and that of $\chi^\dagger\sigma^i\chi\bar u\left(\frac{i}2\epsilon^{ijk}\gamma^j\gamma^k\right)u$ both cancel in the matching. Surprisingly, also the coefficient of the structure $\chi^\dagger\chi\bar u \gamma^0u$  cancels in the matching. We conclude that for the toy example of a point particle,
\begin{equation}\label{b1b2pp}
b_1^{\mbox{\scriptsize p.p.}}=0,\quad b_2^{\mbox{\scriptsize p.p.}}=Q_l^2Z^2\alpha^2\left[\dfrac{16}{3}+\log\left(\dfrac{M}{2\Lambda}\right)\right].
\end{equation} 
One may ask at which order would the  structure $\chi^\dagger\chi\bar u \gamma^0u$ get a non-zero contribution. Looking at equation (\ref{QN_pp_FG}), the only term that is not divergent in the $\lambda\to 0$ or $m\to 0$ limit is $4m\log M/[M(m^2-M^2)]$. This term vanishes at power $1/M^2$, but at power $1/M^3$ it gives $-4m\log M/M^3$. Since when expanding the propagator $\log M$ cannot be generated by the effective field theory calculation, we expect that this would give rise to a non-zero contribution to a matching coefficient.  Since we have not calculated the QED-NRQED amplitude to power $1/M^3$, we cannot determine it at this stage.

We emphasize again that the results in (\ref{b1b2pp}) are for a toy example that in no way represents the behavior of the real proton. We consider the case of the proton next.  

\subsection{Extraction of matching coefficients for the proton}
For the proton the full theory amplitude is given by (\ref{FullSimplified}). To find $b_1$ and $b_2$ we only need the  structures $\chi^\dagger\chi\bar u \gamma^0u$ and $\chi^\dagger\sigma^i\chi\bar u\gamma^i\gamma^5u$. The QED-NRQED result is given in (\ref{Total_EFT}). The Wilson coefficient $b_1$ is determined by the relation
\begin{align}\label {b1extraction}
&\Bigg[Z^2\left(\dfrac{2m \pi}{\lambda^3}-\dfrac{2m^2 \pi}{M\lambda^3}+\dfrac{2m^3 \pi}{M^2\lambda^3}-\dfrac{5 \pi}{4M\lambda}+\dfrac{3 m\pi}{4M^2\lambda}-\dfrac{2}{3mM}+\dfrac{4\log(m/\lambda)}{mM}\right)+c_F^2 \dfrac{m\pi}{M^2\lambda}-c_DZ \dfrac{m\pi}{2M^2\lambda}\Bigg]+\nonumber\\
&+\dfrac{b_1(\alpha^2Q_\ell^2)^{-1}}{M^2}=\nonumber\\
&=\dfrac{2}{\pi}\dfrac{m}{M}\int_0^\infty dQ\,Q^3\int_{-1}^{1}dx\sqrt{1-x^2}\,\dfrac{(1-4x^2)W_1(2iMQx,Q^2)+\left(1-x^2\right)M^2W_2(2iMQx,Q^2)}{\left(Q^2+\lambda^2\right)^2(Q^2+4m^2x^2)}.
\end{align}  
The Wilson coefficient $b_2$ is determined by the relation\footnote{In order to determine $b_2$ we must differentiate between $\bar u\gamma^i\gamma^5u$ and $\bar u(\frac{i}{2}\epsilon^{ijk}\gamma^j\gamma^k)u$. In the non-relativistic limit both terms give $\chi_\ell^\dagger\sigma^i\chi_\ell$. The expression given, e.g., in \cite{Drell:1966kk} corresponds to the non-relativistic limit and not to the expressions we need here. Similar qualifications apply for $b_1$ and the right hand side of (\ref{b1extraction}).}
\begin{align}\label {b2extraction}
&\Bigg[c_FZ \left(-\dfrac{4\pi}{3M\lambda}+\dfrac{2m\pi}{3M^2\lambda}+\dfrac{2\log(2\Lambda/\lambda)}{M^2}+\dfrac{2\log(2\Lambda/m)}{M^2}-\dfrac{16}{3M^2}\right)+\nonumber\\
&+c_F^2 \left(-\dfrac{m\pi}{3M^2\lambda}+\dfrac{\log(m/\lambda)}{M^2}-\dfrac{\log(2\Lambda/m)}{2M^2}-\dfrac{1}{12M^2}\right)+c_SZ \left(-\dfrac{\log(m/\lambda)}{M^2}-\dfrac{3\log(2\Lambda/m)}{2M^2}+\dfrac{13}{12M^2}\right)\Bigg]\nonumber\\
&+\dfrac{b_2(\alpha^2Q_\ell^2)^{-1}}{M^2}=\dfrac{8}{3\pi M}\int_0^\infty dQ\,Q^3\int_{-1}^{1}dx\sqrt{1-x^2}\,\dfrac{1}{\left(Q^2+\lambda^2\right)^2(Q^2+4m^2x^2)}\times\nonumber\\
&\Big[M(2Q^2+x^2Q^2+6m^2x^2)H_1(2iMQx,Q^2)-\left(3ixQ^3+2iQxm^2+2iQx^3m^2\right)H_2(2iMQx,Q^2)\Big].
\end{align}  
Equations (\ref{b1extraction}) and (\ref{b2extraction}) are the main results of this paper.

Given the expressions for $W_1, W_2, H_1, H_2$ one can find an explicit expression for $b_1$ and $b_2$. In lieu of that,  we can use (\ref{Full_expanded}) to find the contributions to $b_1$ and $b_2$ from the full theory that are proportional to $F_1(0)$, $F_2(0)$ and $M^2 F_1^\prime(0)$. Comparing (\ref{Full_expanded}) to (\ref{Total_EFT}) we find first that there is no contribution to matching coefficients for the structure $\chi^\dagger\chi\bar u u$ and $\chi^\dagger\sigma^i\chi\bar u\left(\frac{i}2\epsilon^{ijk}\gamma^j\gamma^k\right)u$ at power $1/M^2$ as expected from chiral symmetry of the leptons and similar to the point particle toy example. The contributions to  $b_1$ and $b_2$ are 
\begin{eqnarray}
b_1(\alpha^2Q_\ell^2)^{-1}&=&0+ \mbox{non } F_1(0), F_2(0), M^2 F_1^\prime(0) \mbox{ terms}\nonumber\\
b_2(\alpha^2Q_\ell^2)^{-1}&=&F_1(0)^2\left[\dfrac{16}{3}+\log\left(\dfrac{M}{2\Lambda}\right)\right]+F_1(0)F_2(0)\cdot \dfrac{16}{3}+\nonumber\\
&+&F_2(0)^2\left[\dfrac{17}{24}-\dfrac12\log\left(\dfrac{M}{2\Lambda}\right)+\dfrac32\log\left(\dfrac{Q}{M}\right)\right]
+ \mbox{non } F_1(0), F_2(0), M^2 F_1^\prime(0) \mbox{ terms.}\nonumber\\
\end{eqnarray}
As discussed in the previous section, the $\log(Q/M)$  is a UV divergence that arises from the Taylor expansion of the form factor. Such a term would be regulated when using the full functional form of $F_2$. The $F_1(0)^2$ term agrees with the toy example of a point particle. 

Surprisingly, there is no contribution to $b_1$. Would that still hold if we include the full expression for $W_1$ and $W_2$? Comparing (\ref{b1extraction}) and (\ref{b2extraction}) we see that the former has an explicit factor of $m$ in the left hand side. This extra factor of $m$ can be understood by looking at (\ref{FullQ}). All the spin-independent terms in (\ref{FullQ}) contain an explicit factor of $m$ with the exception of the term proportional to $2iQx$. Since this term multiplies $W_1$ which is an even function of $x$, we must combine it with $2imQx$ from the lepton propagator. This introduced an extra factor of $m$ also for this term, leading to the overall factor of $m$ in (\ref{b1extraction}).  As a result, the entire spin-independent contribution has an extra factor of $m$. The only difference between the two  structures is the coefficients of $W_1$ in (\ref{FullSimplified}), $-3m$ for  $\chi^\dagger\chi\bar u u$ and $m(1-4x^2)$ for $\chi^\dagger\chi\bar u \gamma^0u$. We see no reason why the coefficient of the former would receive no contribution at power $1/M^2$, while the coefficient of the latter would, when including the full expression for $W_1$.

\section{Conclusions and outlook}\label{sec:conclusions}
Motivated by the proton radius puzzle, a new elastic muon-proton scattering experiment called MUSE will soon start taking data. In this experiment muons with energy close to the muon mass scatter of protons. In such an energy regime one can describe the proton using NRQED, but one has to use QED to describe the muon. An appropriate effective field theory for such kinematics is QED-NRQED.  

QED-NRQED is an effective field theory that describes the electromagnetic interaction of a relativistic point particle, e.g. a lepton, with a non-relativistic spin-half, possibly composite, particle, e.g. a proton. The QED-NRQED Lagrangian can be organized in inverse powers of $M$, the non-relativistic particle mass. For addressing the proton radius puzzle the most important operators are the dimension six spin-independent ones, namely the Darwin term $\psi^\dagger c_De{[\bm{\nabla\cdot E}]}\psi/{8M^2}$, and the contact interaction $b_1\psi^\dagger\psi\,\bar \ell\gamma^0\ell/M^2$. The Wilson coefficient $c_D$ is related to the proton charge radius, and $b_1$ encodes two-photon exchange contributions from physics above the scale $M$. 

In a previous paper we have shown how this effective field theory reproduces the Rosenbluth scattering amplitude at ${\cal O}(Z\alpha)$ and power $1/M^2$ and a relativistic scattering off a static potential at ${\cal O}(Z^2\alpha^2)$ scattering at leading power in $1/M$. In this paper we determined the Wilson coefficients of the contact interactions at ${\cal O}(Z^2\alpha^2)$ and power $1/M^2$. 

To determine the coefficients we did a matching calculation where we match a full theory onto QED-NRQED. We have considered two cases of full theories:  A toy example of a non-relativistic point particle, and the real proton which is described by a hadronic tensor. For both cases we have done the matching calculation in Feynman as well as in Coulomb gauge. 

The effective field theory calculation in Feynman gauge is described in section \ref{sec:QN} and in Coulomb gauge in appendix \ref{app:CG}. We calculated the amplitude for forward off-shell $\ell+p\to\ell+p$ scattering at ${\cal O}(Z^2\alpha^2)$ and power $1/M^2$ in QED-NRQED. The relevant Feynman diagrams are shown in figure \ref{QN_Diagrams}. The typical loop integrals in QED-NRQED involve both relativistic and non-relativistic propagators. We presented two methods for such a calculation and explicitly demonstrated their equivalence for the leading power diagrams. The easier method to use involves an expansion of the NRQED propagator in powers of $1/M$ before performing the loop integrals. The complete QED-NRQED amplitude is given in (\ref{Total_EFT}) for Feynman gauge and in (\ref{Total_EFT_CG}) for Coulomb gauge. The two amplitudes are not the same. There are four structures in the amplitude $\chi^\dagger\chi\bar u u$, $\chi^\dagger\sigma^i\chi\bar u\left(\frac{i}{2}\epsilon^{ijk}\gamma^j\gamma^k\right)u$, $\chi^\dagger\chi\bar u \gamma^0u$, and $\chi^\dagger\sigma^i\chi\bar u\gamma^i\gamma^5u$. The first two involve an even number of Dirac gamma matrices and are expected to be suppressed by the chiral symmetry of the leptons. Only the last two are expected to lead to a non-zero Wilson coefficient at power $1/M^2$. 

The full theory calculation in Feynman gauge is described in section \ref{sec:Full} and in Coulomb gauge in appendix \ref{app:CG}. For the toy example of the non-relativistic point particle we presented an explicit expression for the amplitude in (\ref{QN_pp_FG}) for Feynman gauge and in (\ref{QN_pp_CG}) for Coulomb gauge. As in the effective field theory case, the two amplitudes are different in different gauges. Interestingly when we add in pairs the spin-independent terms and the spin-dependent terms we find that the two gauges agree. We explain the reason in appendix \ref{app:GI}. For the case of a real proton we present an implicit expression for the amplitude given in terms of integrals over the scalar functions that multiply the components of the hadronic tensor. The expression for the amplitude in Feynman gauge is given in (\ref{FullSimplified}) and in Coulomb gauge in (\ref{master_formula_CG}). The two amplitudes are different in the different gauges.

We performed the matching and the extraction of the Wilson coefficients $b_1$ and $b_2$ in section \ref{sec:Extraction}. Both the full and effective field theory amplitudes are IR divergent. We regulate these IR singularities by using a fictitious photon ``mass" denoted by $\lambda$. The lepton mass $m$ is also an IR quantity in the full and the effective theory. Each type of IR singularities, as well as mixed IR singularities, e.g.  $\log (m/\lambda)/(mM)$, cancel in the matching.  For the toy example of a non-relativistic point particle the explicit expressions for $b_1$ and $b_2$ are given in (\ref{b1b2pp}) for Feynman gauge and in (\ref{b1b2pp_CG}) for Coulomb gauge. Although both the full and effective theory amplitudes are different in each gauge, the Wilson coefficients are the same for both gauges. Surprisingly we find that $b_1^{\mbox{\scriptsize p.p.}}=0$ at ${\cal O}(Z^2\alpha^2)$. For the case of the proton we give implicit expressions for $b_1$ and $b_2$ in (\ref{b1extraction}) and (\ref{b2extraction}) for Feynman gauge. These are the main results of the paper. We show explicitly how the IR singularities cancel between the full and effective theory in both Feynman and Coulomb gauge based on the known low-energy behavior of the hadronic tensor. We find that $b_1=0$ at ${\cal O}(Z^2\alpha^2)$. Based on the structure of the full theory integral in Coulomb gauge (\ref{FullSimplified}) and in Feynman gauge (\ref{master_formula_CG}) we argued that $b_1=0$  at ${\cal O}(Z^2\alpha^2)$ even if we include the full expression of the hadronic tensor. The reason is that the full theory expression includes an explicit factor of $m$ in those integrals for all the spin-independent terms. The spin-independent part of the effective field theory expression contains only IR singular terms that cancel in the matching and do not contribute to $b_1$. 

The fact that $b_1=0$ at ${\cal O}(Z^2\alpha^2)$ is surprising. It does not obviously follow from a symmetry. One might ask at which power two-photon exchange effects from scales above $M$ enter. For the toy example of a point particle there is a term in the full theory amplitude that would give rise to such a spin-independent matching coefficient at power $1/M^3$. At this power we expect to find also other operators, some of which depend on the lepton momentum.  

This implies that scattering amplitudes calculated within QED-NRQED are not sensitive to two-photon exchange effects from scales above $M$ at  ${\cal O}(Z^2\alpha^2)$ and power $1/M^2$. At power $1/M^2$ the spin-independent part of the amplitude depends only on $c_D$, or in other words, on the proton charge radius.  For the case of $m$ being the muon mass and $M$ the proton mass, we conclude that two-photon exchange effects from scales above $M$ are an order of magnitude smaller compared to proton charge radius effects. One of the suggested explanations for the proton radius puzzle involves an unusual behavior of $W_1(0,Q^2)$, since only its asymptotic low and high $Q^2$ terms are known. The vanishing of $b_1$  at ${\cal O}(Z^2\alpha^2)$ implies that the MUSE experiment will be less sensitive to such effects, but its extraction of the proton charge radius will be more robust.   

The vanishing of $b_1$  at ${\cal O}(Z^2\alpha^2)$ arises from a combination of two phenomena. The first arises since the effective field theory amplitude contains only IR divergent terms for the spin-independent part of the amplitude. This happens since it is a sum of direct and crossed integrals that tend to cancel each other. Such IR terms must vanish in the matching and give no contribution to $b_1$. The second effect arises from the full theory amplitude. The spin-independent parts of the hadronic tensor, namely $W_1$ and $W_2$,  are multiplied by an explicit factor of $m$ or a term linear in the photon energy (measured in the proton rest frame). Since $W_1$ and $W_2$ are even functions of the photon energy such terms vanish unless they are multiplied by a product of the lepton mass and the photon energy from the lepton propagator, see (\ref{master_formula}) for Feynman gauge and (\ref{master_formula_CG}) for Coulomb gauge. The combination of the two phenomena leads to the vanishing of $b_1$  at ${\cal O}(Z^2\alpha^2)$. 

The dependence of the full theory amplitude on the lepton mass was noted before in \cite{Pineda:2004mx}. This fact was used to argue that the Wilson coefficient $c_{3,R}^{pl_i}$ \cite{Pineda:2002as} of the HBET operator, analogous to $b_1$, is suppressed by an extra factor $m_{l_i} /m_p$, i.e. $c_{3,R}^{pl_i}\sim\alpha^2m_{l_i} /m_p$.  An explicit matching calculation was not done in those papers. The explicit matching calculation of QCD onto QED-NRQED performed in this paper, confirms that indeed the Wilson coefficient vanishes  at ${\cal O}(Z^2\alpha^2)$. 

Beyond the proton radius puzzle, this result can be of interest in physics beyond the standard model, where generating hierarchies, even ``little" ones, between the weak scale and the scale of new physics is an active topic of research. It would be interesting to see if the vanishing of $b_1$ at ${\cal O}(Z^2\alpha^2)$  can be used  to generate such hierarchies. 

\vskip 0.2in
\noindent
{\bf Acknowledgements}
\vskip 0.1in
\noindent
We  thank Andrew E. Blechman and Alexey A. Petrov for useful discussions and comments on the manuscript. We also thank Roni Harnik, Richard J. Hill, Andreas S. Kronfeld, Antonio Pineda, and Randolf Pohl for useful discussions. This work was supported by the U.S. Department of Energy grant DE-SC0007983, by a grant from the Simons Foundation (562836, G.P.), and by Fermilab's Intensity Frontier Fellowship. G.P. thanks Fermilab and KITP for their hospitality and support during the completion of this work. KITP is supported in part by the National Science Foundation under Grant No. NSF PHY17-48958. This manuscript has been authored by Fermi Research Alliance, LLC, under Contract No. DE-AC02-07CH11359 with the U.S. Department of Energy, Office of Science, Office of High Energy Physics.

\begin{appendix}
\section{Appendix: Matching in Coulomb gauge}\label{app:CG}
\subsection{Matching in Coulomb gauge}
The Coulomb gauge is often used for NRQED calculations \cite{Kinoshita:1995mt}. Apart from Feynman gauge we have performed the matching also in Coulomb gauge. The QED-NRQED amplitude in  Coulomb gauge is 
\begin{align}\label{Total_EFT_CG}
\dfrac{{\cal M}^{\mbox{\scriptsize EFT}}}{\alpha^2Q_\ell^2}&=\chi^\dagger\chi\bar u u\Bigg[Z^2\left(\dfrac{2m \pi}{\lambda^3}-\dfrac{2m^2 \pi}{M\lambda^3}+\dfrac{2m^3 \pi}{M^2\lambda^3}+\dfrac{17 \pi}{16M\lambda}+\dfrac{7 m\pi}{16M^2\lambda}-\dfrac{247}{105mM}-\dfrac{2\log(m/\lambda)}{mM}\right)\nonumber\\
&\qquad\qquad\quad{}-c_F^2 \dfrac{m\pi}{M^2\lambda}-c_DZ \dfrac{m\pi}{2M^2\lambda}\Bigg]+\nonumber\\
&+\chi^\dagger\chi\bar u\gamma^0 u\Bigg[Z^2\left(\dfrac{2m \pi}{\lambda^3}-\dfrac{2m^2 \pi}{M\lambda^3}+\dfrac{2m^3 \pi}{M^2\lambda^3}-\dfrac{25 \pi}{16M\lambda}+\dfrac{17 m\pi}{16M^2\lambda}+\dfrac{107}{105mM}+\dfrac{4\log(m/\lambda)}{mM}\right)\nonumber\\
&\qquad\qquad\qquad{}+c_F^2 \dfrac{m\pi}{M^2\lambda}-c_DZ \dfrac{m\pi}{2M^2\lambda}\Bigg]+\nonumber\\
&+\chi^\dagger\sigma^i\chi\bar u\left(\frac{i}2\epsilon^{ijk}\gamma^j\gamma^k\right)u\Bigg[c_FZ \left(\dfrac{m\pi}{6M^2\lambda}-\dfrac{2}{5M^2}\right)+c_F^2 \left(\dfrac{m\pi}{3M^2\lambda}-\dfrac{\log(m/\lambda)}{M^2}\right)\nonumber\\
&\qquad\qquad\qquad\qquad\qquad\quad\,{}+c_SZ \left(-\dfrac1{9M^2}+\dfrac{\log(m/\lambda)}{3M^2}\right)\Bigg]+\nonumber\\
&+\chi^\dagger\sigma^i\chi\bar u\gamma^i\gamma^5u\Bigg[c_FZ \left(-\dfrac{4\pi}{3M\lambda}+\dfrac{7m\pi}{6M^2\lambda}+\dfrac{4\log(2\Lambda/m)}{M^2}-\dfrac{74}{15M^2}\right)\nonumber\\
&\qquad\qquad\qquad\quad{}+c_F^2 \left(-\dfrac{m\pi}{3M^2\lambda}+\dfrac{\log(m/\lambda)}{M^2}-\dfrac{\log(2\Lambda/m)}{2M^2}-\dfrac{1}{12M^2}\right)\nonumber\\
&\qquad\qquad\qquad\quad{}+c_SZ \left(-\dfrac{\log(m/\lambda)}{3M^2}-\dfrac{3\log(2\Lambda/m)}{2M^2}+\dfrac{7}{36M^2}\right)\Bigg].
\end{align}  
Comparing to the QED-NRQED result in Feynman gauge (\ref{Total_EFT}) we notice several differences in each structure. Interestingly, if we add the terms multiplying $\chi^\dagger\chi\bar u u$ and $\chi^\dagger\chi\bar u\gamma^0 u$ we find the same answer as the corresponding sum in (\ref{Total_EFT}). The same is true if we add the terms multiplying $\chi^\dagger\sigma^i\chi\bar u\left(\frac{i}2\epsilon^{ijk}\gamma^j\gamma^k\right)u$ and $\chi^\dagger\sigma^i\chi\bar u\gamma^i\gamma^5u$. 

We turn now to the full theory calculation. For the toy example of a non-relativistic point particle we find  in Coulomb gauge
\begin{eqnarray}\label{QN_pp_CG} 
&&\dfrac{{\cal M}^{\scriptsize\mbox{p.p.}}}{Q_l^2Z^2\alpha^2}=\chi^\dagger\chi\bar uu\left[\dfrac{2mM\pi}{(m+M)\lambda^3}+\dfrac{17\pi}{16(m+M)\lambda}+\dfrac1{mM}\left(2\log \lambda-\dfrac{247}{105}-\dfrac{2\left(m^2\log M-M^2\log m\right)}{m^2-M^2}\right)\right]\nonumber\\
&+&\chi^\dagger\chi\bar u\gamma^0u\left[\dfrac{2mM\pi}{(m+M)\lambda^3}-\dfrac{25\pi}{16(m+M)\lambda}+\dfrac1{mM}\left(-4\log \lambda+\dfrac{107}{105}+\dfrac{4(m^2\log M-M^2\log m)}{m^2-M^2}\right)\right]\nonumber\\
&+&\chi^\dagger\sigma^i\chi\bar u\left(\frac{i}2\epsilon^{ijk}\gamma^j\gamma^k\right)u\left[\dfrac{m\pi}{2M(m+M)\lambda}+\dfrac{2\log\lambda}{3M^2}-\dfrac{23}{45M^2}+\dfrac{2\left(M^2\log m-m^2\log M\right)}{3M^2(m^2-M^2)}\right]\nonumber\\
&+&\chi^\dagger\sigma^i\chi\bar u\gamma^i\gamma^5u
\left[-\dfrac{(3m+8M)\pi}{6M(m+M)\lambda}-\dfrac{2\log\lambda}{3M^2}+\dfrac{23}{45M^2}+\dfrac{4\left(M^2\log m-m^2\log M\right)}{3M^2(m^2-M^2)}+\dfrac{2\log M}{M^2}\right].
\end{eqnarray}
Again the result is different from (\ref{QN_pp_FG}). If we add in pairs the spin-independent terms and the spin-dependent terms we find agreement with (\ref{NN_pp_FG}). We explain this result in appendix \ref{app:GI}.

Next we look at the  proton amplitude  in terms of the hadronic tensor. Since the photon propagator is different for time-like and space-like indices, we get a more complicated result compared to the Feynman gauge amplitude (\ref{master_formula}):
\begin{eqnarray}\label{master_formula_CG}
i{\cal M}&=&Q_\ell^2e^4\int\dfrac{d^4l}{(2\pi)^4}\dfrac{1}{l^2-2ml^0}\,\dfrac1{2M}\times\Bigg\{2\chi^\dagger\chi \bar u\gamma^0u\dfrac{\vec{l}^{\,2}\lambda^2}{(\vec{l}^{\,2}+\lambda^2)^2}\dfrac1{l^2-\lambda^2}\frac{l^0}{l^2}\left[W_1+M^2\frac{\vec{l}^{\,\,2}}{l^2}W_2\right]
\nonumber\\
&&-\chi^\dagger\chi \left[\bar u\gamma^0u (m-l^0)+m \bar uu\right]\left(\dfrac{1}{\vec{l}^{\,2}+\lambda^2}\right)^2\frac{\vec{l}^{\,\,2}}{l^2}\left[W_1+M^2\frac{\vec{l}^{\,\,2}}{l^2}W_2\right]
\nonumber\\
&&+\chi^\dagger\chi \left[\bar u\gamma^0u (l^0-m)+m \bar uu\right]\left(\dfrac{1}{l^2-\lambda^2}\right)^2\left[W_1\left(2+\dfrac{\lambda^4(l^0)^2}{l^2(\vec{l}^{\,2}+\lambda^2)^2}\right)+M^2\frac{\vec{l}^{\,\,2}}{l^2}\dfrac{\lambda^4(l^0)^2}{l^2(\vec{l}^{\,2}+\lambda^2)^2}W_2\right]\nonumber\\
&&+\chi^\dagger\sigma^i\chi\bar u\left(\frac{i}2\epsilon^{ijk}\gamma^j\gamma^k\right)u\, 4m\left(\dfrac1{l^2-\lambda^2} \right)^2\left[\left(Ml^0H_1+l^2H_2\right)\left(1-\dfrac{2\vec{l}^{\,2}}{3(\vec{l}^{\,2}+\lambda^2)}\right)+H_2\dfrac{2\vec{l}^{\,2}\lambda^2}{3(\vec{l}^{\,2}+\lambda^2)}\right]\nonumber\\
&&+\chi^\dagger\sigma^i\chi\bar u\gamma^i\gamma^5u(l^0-m)\, 4m\left(\dfrac1{l^2-\lambda^2} \right)^2\left[4\left(Ml^0H_1+l^2H_2\right)\left(1-\dfrac{2\vec{l}^{\,2}}{3(\vec{l}^{\,2}+\lambda^2)}\right)+H_2\dfrac{8\vec{l}^{\,2}\lambda^2}{3(\vec{l}^{\,2}+\lambda^2)}\right]\nonumber\\
&&+\chi^\dagger\sigma^i\chi\bar u\gamma^i\gamma^5u\,\dfrac{8\vec{l}^{\,\,2}}{3(\vec{l}^{\,2}+\lambda^2)(l^2-\lambda^2)}\left[MH_1+l^0H_2\right]\Bigg\}.
\end{eqnarray}
We now change variables as in the case of Feynman gauge. Inserting the expressions from (\ref{Wsing}), performing the integrals, and expanding in inverse powers of $M$ gives 
\begin{align}\label{Full_expanded_CG}
&\dfrac{{\cal M}^{\mbox{\scriptsize Full}}_{\mbox{\scriptsize Expanded}}}{\alpha^2Q_\ell^2}=\chi^\dagger\chi\bar u u\Bigg[F_1(0)^2\left(\dfrac{2m \pi}{\lambda^3}-\dfrac{2m^2 \pi}{M\lambda^3}+\dfrac{2m^3 \pi}{M^2\lambda^3}+\dfrac{17 \pi}{16M\lambda}-\dfrac{17 m\pi}{16M^2\lambda}-\dfrac{247}{105mM}-\dfrac{2\log(m/\lambda)}{mM}\right)\nonumber\\
&\hspace{9.5em}-F_1(0)F_2(0) \dfrac{3m\pi}{M^2\lambda}-F_2(0)^2 \dfrac{m\pi}{M^2\lambda}-F_1(0)M^2 F_1^\prime(0)\dfrac{4m\pi}{M^2\lambda}\Bigg]+\nonumber\\
&+\chi^\dagger\chi\bar u \gamma^0u\Bigg[F_1(0)^2\left(\dfrac{2m \pi}{\lambda^3}-\dfrac{2m^2 \pi}{M\lambda^3}+\dfrac{2m^3 \pi}{M^2\lambda^3}-\dfrac{25 \pi}{16M\lambda}+\dfrac{25 m\pi}{16M^2\lambda}+\dfrac{107}{105mM}+\dfrac{4\log(m/\lambda)}{mM}\right)\nonumber\\
&\qquad\qquad\quad{}+F_1(0)F_2(0) \dfrac{m\pi}{M^2\lambda}+F_2(0)^2 \dfrac{m\pi}{M^2\lambda}-F_1(0)M^2 F_1^\prime(0)\dfrac{4m\pi}{M^2\lambda}\Bigg]+\nonumber\\
&+\chi^\dagger\sigma^i\chi\bar u\left(\frac{i}2\epsilon^{ijk}\gamma^j\gamma^k\right)u\Bigg[F_1(0)^2\left(\dfrac{m\pi}{2M^2\lambda}-\dfrac{2\log(m/\lambda)}{3M^2}-\dfrac{23}{45M^2}\right)+\nonumber\\
&\hspace{10.5em}+F_1(0)F_2(0) \left(\dfrac{5m\pi}{6M^2\lambda}-\dfrac{4\log(m/\lambda)}{3M^2}-\dfrac{28}{45M^2}\right)+\nonumber\\
&\hspace{10.5em}+F_2(0)^2 \left(\dfrac{m\pi}{3M^2\lambda}-\dfrac{\log(m/\lambda)}{M^2}\right)\Bigg]+\nonumber\\
&+\chi^\dagger\sigma^i\chi\bar u\gamma^i\gamma^5u\Bigg[F_1(0)^2\left(-\dfrac{4\pi}{3M\lambda}+\dfrac{5m\pi}{6M^2\lambda}+\dfrac{2\log(M/\lambda)}{3M^2}+\dfrac{4\log(M/m)}{3M^2}+\dfrac{23}{45M^2}\right)+\nonumber\\
&\hspace{6.5em}+F_2(0)^2 \left(-\dfrac{m\pi}{3M^2\lambda}+\dfrac{\log(m/\lambda)}{M^2}-\dfrac{\log(M/m)}{2M^2}+\dfrac{3\log(Q/M)}{2M^2}+\dfrac{5}{8M^2}\right)\Bigg] \nonumber\\
&\hspace{6.5em}+F_1(0)F_2(0) \left(-\dfrac{4\pi}{3M\lambda}+\dfrac{m\pi}{2M^2\lambda}+\dfrac{4\log(m/\lambda)}{3M^2}+\dfrac{28}{45M^2}\right)\Bigg]+{\cal O}\left(\frac1{M^3}\right).
\end{align}

As for Feynman gauge, the $\log(Q/M)$ in the last line is a UV divergence that arises from the Taylor expansion of the form factor. Such a term would be regulated when using the full functional form of $F_2$. 

As a check, we can add the $\chi^\dagger\chi\bar u u$ and $\chi^\dagger\chi\bar u\gamma^0 u$ terms and compare the sum to the NRQED result on the left hand side of equation (7) of \cite{Hill:2011wy}. Setting $F_1(0)=1$ and expanding that result to order $1/M^2$, we find a complete agreement.  As another check, the terms in (\ref{Full_expanded_CG}) proportional to $F_1(0)^2$ match the IR singular terms of (\ref{QN_pp_CG}) expanded to order $1/M^2$.

We can use these results to extract the matching coefficients following the same procedure as in section \ref{sec:Extraction}.  First, for the toy example of a non-relativistic point particle. Setting the Wilson coefficients to their point particle values in (\ref{Total_EFT_CG}) and subtracting the result from the order $1/M^2$ expansion of (\ref{QN_pp_CG}) we find that in Coulomb gauge  
\begin{equation}\label{b1b2pp_CG}
b_1^{\mbox{\scriptsize p.p.}}=0,\quad b_2^{\mbox{\scriptsize p.p.}}=Q_l^2Z^2\alpha^2\left[\dfrac{16}{3}+\log\left(\dfrac{M}{2\Lambda}\right)\right].
\end{equation} 
This is the same result as in Feynman gauge, (\ref{b1b2pp}). Although the full and effective field theory both differ between Feynman and Coulomb gauge, the Wilson coefficients are the same in both gauges.  

For the case of the proton the Coulomb gauge expressions analogous to (\ref{b1extraction}) and  (\ref{b2extraction}) are longer and we will not give them here. To compare to Feynman gauge, we find the contributions to $b_1$ and $b_2$ from the full theory that are proportional to $F_1(0)$, $F_2(0)$ and $M^2 F_1^\prime(0)$. These are 
\begin{eqnarray}
b_1(\alpha^2Q_\ell^2)^{-1}&=&0+ \mbox{non } F_1(0), F_2(0), M^2 F_1^\prime(0) \mbox{ terms}\nonumber\\
b_2(\alpha^2Q_\ell^2)^{-1}&=&F_1(0)^2\left[\dfrac{16}{3}+\log\left(\dfrac{M}{2\Lambda}\right)\right]+F_1(0)F_2(0)\cdot \dfrac{16}{3}+\nonumber\\
&+&F_2(0)^2\left[\dfrac{17}{24}-\dfrac12\log\left(\dfrac{M}{2\Lambda}\right)+\dfrac32\log\left(\dfrac{Q}{M}\right)\right]
+ \mbox{non } F_1(0), F_2(0), M^2 F_1^\prime(0) \mbox{ terms.}\nonumber\\
\end{eqnarray}
This is the same result as in Feynman gauge, (\ref{b1b2pp}). In particular there is no contribution to $b_1$. Since in  (\ref{master_formula_CG}) the spin independent terms either include an explicit factor of $m$ or a term linear in $l^0$, we expect no contribution to $b_1$ even if we include the full expression for $W_1$ and $W_2$.

\subsection{Gauge invariance}\label{app:GI}
We have found that the matching coefficients are equal for Feynman and Coulomb gauges. The effective and full theory amplitudes, on the other hand, are different in each gauge. This can be expected.  Since we want to distinguish, e.g.,  $\bar u\gamma^0 u$ from $\bar u  u$, we cannot put the lepton on-shell. Since the amplitudes are off-shell, they are not guaranteed to be gauge independent. This is most prominent for the point particle case where we can compute the full theory amplitude explicitly. Surprisingly, when taking the non-relativistic limit for the lepton, the two gauges agree. We now explain this for the non-relativistic point particle case.   

For a non-relativistic point particle case we take $u(p)\to(\chi\,\, 0)^T$. At its rest frame, $p^0=M$,  $(\pslash-M)u(p)\to(p^0\gamma^0-M)(\chi\,\, 0)^T=0$. In other words, the non-relativistic particle is ``effectively" on-shell. Consider now a general covariant gauge where the photon propagator is 
\begin{equation}
D^{\mu\nu}(l)=\dfrac{-i}{l^2-\lambda^2}\left[g^{\mu\nu}+(\xi-1)\dfrac{l^\mu l^\nu}{l^2-\lambda^2}\right].
\end{equation} 
The $(\xi-1)$ terms add an $\lslash$ on the non-relativistic point particle line. For example, one of the terms gives $\bar u(p)\gamma^\mu\left[\lslash+M(1+\gamma^0)\right]\lslash u(p)$ in the general covariant gauge analog of (\ref{AB}). We now write $\lslash =\pm(\pslash\pm\lslash-M) \mp(\pslash-M)$. The first term cancels the non-relativistic point particle propagator and will vanish when we add the direct and crossed contribution to the amplitude. The second term vanishes as a result of the ``effective" on-shell condition for the non-relativistic point particle. We conclude that a general covariant gauge amplitude is the same as that of a Feynman gauge.

We can try and apply a similar procedure for Coulomb gauge. It turns out that the gauge dependent parts in Coulomb gauge arise from terms that include $\gamma^i l^i$. We can use the identity $\gamma^i l^i=\gamma^0l^0-\lslash$ to generate terms analogous to covariant gauge. But unlike covariant gauge we have terms, e.g.,  $\bar u(k) \gamma^i l^i \dfrac1{ \kslash+\lslash-m} \gamma ^0 u(k)\bar u(p)  \gamma^j\l^j \dfrac1{\pslash-\lslash-M} \gamma ^0 u(p)$. Using  $\gamma^i l^i=\gamma^0l^0-\lslash$ generates terms with an $\lslash$ only on the lepton line. Such terms do not vanish for a relativistic lepton thus generating Coulomb gauge dependent pieces.  Once we take the lepton to be also non-relativistic, we can apply the same procedure as before and use that $\kslash-m$ annihilates the lepton spinor in the non-relativistic limit.  This explains why in the point particle case the relativistic-non-relativistic amplitude is different in Coulomb gauge while the non-relativistic-non-relativistic amplitude is the same in Coulomb and Feynman gauges.  

We expect a similar reasoning to apply for the QED-NRQED amplitude and the proton amplitude, but we have not proven it in these cases. The former since it must reproduce the IR singularities of the point particle amplitude, and the latter since it is a generalization of the point particle case.

\section{Appendix: Properties of the Hadronic tensor }\label{app:HT}
For completeness we derive several properties of the hadronic tensor $W^{\mu\nu}$ defined in (\ref{Wdefined}). The structure of $W^{\mu\nu}$ is constrained by current conservation, by being a forward matrix element, and by the parity and time reversal symmetries of the electromagnetic and strong interactions.  First, current conservation implies that $q_\mu W^{\mu\nu}=0$ and $q_\nu W^{\mu\nu}=0$. 

Second, $W^{\mu\nu}$ is a forward matrix element so we can write it as
\begin{equation}
W^{\mu\nu}(p,q)=\dfrac1{2\mproton}\bar u_p(p,s) \Gamma^{\mu\nu}(p,q) u_p(p,s),
\end{equation}
where $\Gamma^{\mu\nu}$ is a general Dirac structure. 

The Dirac structure can be simplified by using that for on-shell spinors $\vslash u(p)=u(p)$ where $\vslash\equiv\pslash/m$. In analogy to Heavy Quark Effective Theory (HQET) we can define the projector $P_+\equiv(1+\vslash)/2$, where $P_+u(p)=u(p)$. As was shown in \cite{Mannel:1994kv},  between two $P_+$'s the Dirac basis reduces to four matrices: $P_+$ and $s^\mu=P_+\gamma^\mu\gamma^5P_+$. The matrices $s^\mu$ are a generalization of the Pauli spin matrices that satisfy $v\cdot s=0$. 

To avoid introducing a $\gamma^5$ for parity even theories, we can define 
\begin{equation}
s^{\alpha\beta}\equiv P_+\left(\sigma^{\alpha\beta}-\dfrac{\sigma^{\rho\beta}p^\alpha p_\rho}{M^2}-\dfrac{\sigma^{\alpha\rho}p^\beta p_\rho}{M^2}\right)P_+,
\end{equation} 
where $\sigma^{\alpha\beta}=i\left[\gamma^\alpha,\gamma^\beta\right]/2$. It follows that $p_\alpha s^{\alpha\beta}=0$ and $p_\beta s^{\alpha\beta}=0$.  The relation of $s^{\alpha\beta}$ to $s^\mu$ of \cite{Mannel:1994kv} is $s^\mu=-p_\rho\epsilon^{\rho\alpha\beta\mu}s_{\alpha\beta}/2M$, using the convention $\epsilon_{0123}=1$.  For a general Dirac matrix $\Gamma$ we have 
\begin{equation}
P_+\Gamma P_+=\dfrac12\mbox{Tr}\left[P_+\Gamma\right]+\dfrac14 s^{\alpha\beta}\mbox{Tr}\left[P_+ s_{\alpha\beta}P_+\Gamma\right].
\end{equation}
We thus need to consider only the unit matrix and $s^{\alpha\beta}$. 

Third, the electromagnetic and strong interactions are invariant under parity and time reversal. Under their combined operation, $PT$, 
\begin{equation}
\langle p,s|O(x)|p,s \rangle \stackrel{PT}{\to}\langle p,-s|O^\dagger(-x)|p,-s \rangle.
\end{equation}
Since $\left(J^\mu_{\mbox{\scriptsize e.m.}}\right)^\dagger=J^\mu_{\mbox{\scriptsize e.m.}}$,  
\begin{eqnarray}
&&W^{\mu\nu}(p,q)=i\int d^4x\, e^{iqx}\langle p,s|T\left\{J^\mu_{\mbox{\scriptsize e.m.}}(x)J_{\mbox{\scriptsize e.m.}}^\nu(0)\right\}|p,s \rangle=\nonumber\\
&=&i\int d^4x\, e^{iqx}\langle p,s|\theta(x^0)J^\mu_{\mbox{\scriptsize e.m.}}(x)J^\nu_{\mbox{\scriptsize e.m.}}(0)+\theta(-x^0)J^\nu_{\mbox{\scriptsize e.m.}}(0)J^\mu_{\mbox{\scriptsize e.m.}}(x)|p,s \rangle\nonumber\\
& \stackrel{PT}{\to}&i\int d^4x\, e^{iqx}\langle p,-s|\theta(x^0)J^\nu_{\mbox{\scriptsize e.m.}}(0)J^\mu_{\mbox{\scriptsize e.m.}}(-x)+\theta(-x^0)J^\mu_{\mbox{\scriptsize e.m.}}(-x)J^\nu_{\mbox{\scriptsize e.m.}}(0)|p,-s \rangle\nonumber\\
&=&i\int d^4x\, e^{iqx}\langle p,-s|\theta(x^0)J^\nu_{\mbox{\scriptsize e.m.}}(x)J^\mu_{\mbox{\scriptsize e.m.}}(0)+\theta(-x^0)J^\mu_{\mbox{\scriptsize e.m.}}(0)J^\nu_{\mbox{\scriptsize e.m.}}(x)|p,-s \rangle,
\end{eqnarray}
where in the last line we have used the translation invariance of a forward matrix element.  This implies
\begin{equation}
W^{\mu\nu}(p,q)=\dfrac1{2\mproton}\bar u_p(p,s) \Gamma^{\mu\nu}(p,q) u_p(p,s)=\dfrac1{2\mproton}\bar u_p(p,-s) \Gamma^{\nu\mu}(p,q) u_p(p,-s).
\end{equation}
Since $\bar u_p(p,s)u_p(p,s)$ is independent of the spin, the tensor multiplying the unit matrix in $W^{\mu\nu}$ must be symmetric under the interchange  $\mu\leftrightarrow\nu$. The tensor multiplying $s^{\alpha\beta}$ in $W^{\mu\nu}$ is antisymmetric under the interchange $\mu\leftrightarrow\nu$. This is most easily seen in the rest frame of the proton, where $s^{\alpha\beta}$ are just the Pauli matrices, i.e. $s^{ij}=\epsilon^{ijk}\sigma^k$.  For the Pauli matrices $\chi^\dagger_\uparrow\sigma^i\chi_\uparrow=-\chi^\dagger_\downarrow\sigma^i\chi_\downarrow$

Consider now the identity matrix. Since it is symmetric under $\mu\leftrightarrow\nu$, it is multiplied by a linear combination of the tensors $g^{\mu\nu},p^\mu p^\nu,q^\mu q^\nu$, and $(p^\mu q^\nu+q^\mu p^\nu)$. Using $q_\mu W^{\mu\nu}=0$ and $q_\nu W^{\mu\nu}=0$, we can reduce these to two linear combinations $-g^{\mu\nu}+q^\mu q^\nu/q^2$ and $(p^\mu-p\cdot q\,q^\mu/q^2)(p^\nu-p\cdot q\,q^\nu/q^2)$. 

Consider now $s^{\alpha\beta}$. Instead of using $s^{\alpha\beta}$, we can use $\sigma^{\alpha\beta}$ but omit its contraction with $p^\alpha$ or $p^\beta$. Since it is anti-symmetric, it appears as a linear combination of  $[\gamma^\mu,\gamma^\nu]$, $[\gamma^\nu,\qslash]p^\mu-[\gamma^\mu,\qslash]p^\nu$, and $[\gamma^\nu,\qslash]q^\mu-[\gamma^\mu,\qslash]q^\nu$. Using $q_\mu W^{\mu\nu}=0$ or $q_\nu W^{\mu\nu}=0$ reduces these to two linear combinations, $[\gamma^\nu,\qslash]\,p^\mu- [\gamma^\mu,\qslash]\,p^\nu+[\gamma^\mu,\gamma^\nu]\,p\cdot q$ and $[\gamma^\nu,\qslash]\,q^\mu- [\gamma^\mu,\qslash]\,q^\nu+[\gamma^\mu,\gamma^\nu]\,q^2$.

All together we find equation (\ref{Tensordecomposition}). Since $p^2=M^2$, the scalar coefficients of these structures, $W_1,W_2,H_1,H_2$ depend on $q^2$ and $p\cdot q$, or alternatively on $Q^2$ and $\nu$. Translation invariance of the forward matrix elements implies 
\begin{eqnarray}
&&W^{\mu\nu}(p,q)=i\int d^4x\, e^{iqx}\langle p,s|T\left\{J^\mu_{\mbox{\scriptsize e.m.}}(x)J_{\mbox{\scriptsize e.m.}}^\nu(0)\right\}|p,s \rangle=\nonumber\\
&=&i\int d^4x\, e^{iqx}\langle p,s|\theta(x^0)J^\mu_{\mbox{\scriptsize e.m.}}(x)J^\nu_{\mbox{\scriptsize e.m.}}(0)+\theta(-x^0)J^\nu_{\mbox{\scriptsize e.m.}}(0)J^\mu_{\mbox{\scriptsize e.m.}}(x)|p,s \rangle\nonumber\\
&=&i\int d^4x\, e^{iqx}\langle p,s|\theta(x^0)J^\mu_{\mbox{\scriptsize e.m.}}(0)J^\nu_{\mbox{\scriptsize e.m.}}(-x)+\theta(-x^0)J^\nu_{\mbox{\scriptsize e.m.}}(-x)J^\mu_{\mbox{\scriptsize e.m.}}(0)|p,s \rangle\nonumber\\
&\stackrel{x\to\, -x}{=}&i\int d^4x\, e^{-iqx}\langle p,s|\theta(-x^0)J^\mu_{\mbox{\scriptsize e.m.}}(0)J^\nu_{\mbox{\scriptsize e.m.}}(x)+\theta(x^0)J^\nu_{\mbox{\scriptsize e.m.}}(x)J^\mu_{\mbox{\scriptsize e.m.}}(x)|p,s \rangle\nonumber\\
&=&i\int d^4x\, e^{-iqx}\langle p,s|T\left\{J^\nu_{\mbox{\scriptsize e.m.}}(x)J_{\mbox{\scriptsize e.m.}}^\mu(0)\right\}|p,s \rangle=W^{\nu\mu}(p,-q).
\end{eqnarray}
Combining $W^{\mu\nu}(p,q)=W^{\nu\mu}(p,-q)$ with (\ref{Tensordecomposition}) implies that $W_1,W_2,H_1$ are even functions of $\nu$, while $H_2$ is an odd function of $\nu$.
\end{appendix} 
\section{NRQED Feynman rules}\label{app:FR}
For QED-NRQED amplitudes we need the Feynman rules of both QED and NRQED. The  QED Feynman rules are well-known and will not be listed here. 
Most of the NRQED Feynman rules used in this paper, but not all, are given  in figure 3 of \cite{Kinoshita:1995mt} by multiplying the vertices by $-i$ and the propagators by $i$. That table does not include one of the Feynman rules arising from the term multiplying $c_S$. Also, Coulomb gauge was used in \cite{Kinoshita:1995mt}. In a non-Coulomb gauge, apart from the different photon propagator, there is also an additional interaction from the term multiplying $c_D$. We list below all the Feynman rules up to order $1/M^2$ for both Feynman and Coulomb gauges. 
\subsection{Propagators}
\begin{itemize}
\item The Feynman rule for an NRQED fermion propagator is 
$$
\vcenter{\includegraphics[scale=1]{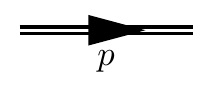}}
\hspace{-30em}   \dfrac{i}{p^{0}-\dfrac{\vec{p}^{\,\,2}}{2M}+i\epsilon}
$$
\item In Feynman gauge the Feynman rule for the photon propagator  is 
$$
\vcenter{\includegraphics[scale=1]{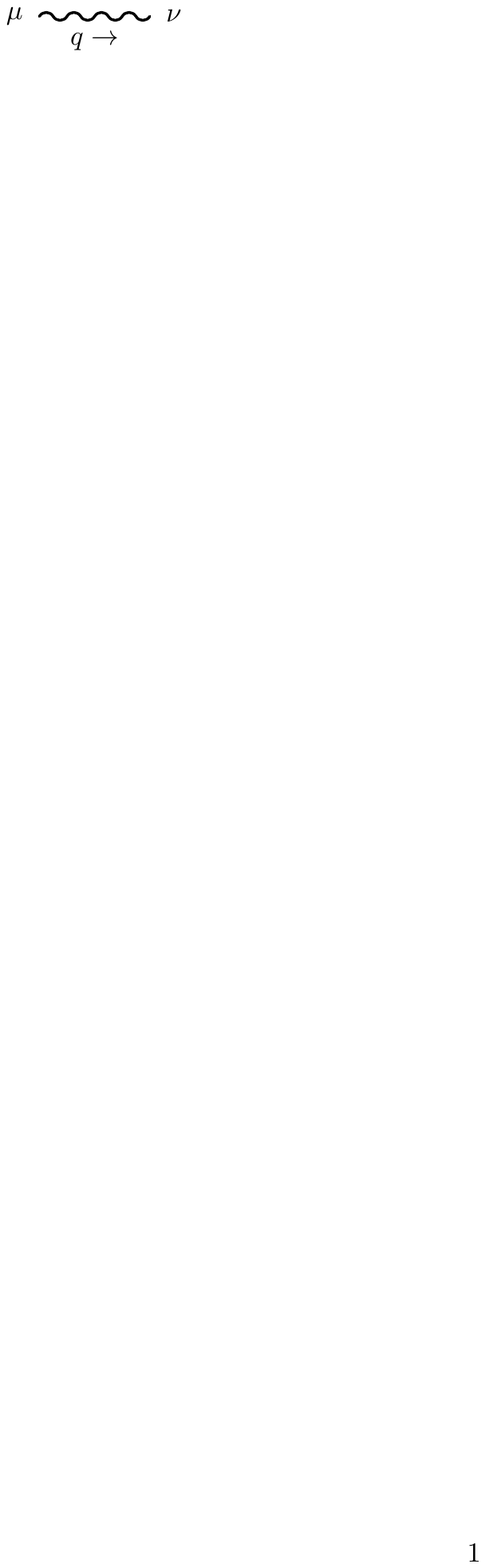}}
\hspace{-30em}   \dfrac{-ig^{\mu \nu}}{q^{2}-\lambda^{2}+i\epsilon}
$$
\end{itemize}
In Coulomb gauge the photon propagator is \cite{Kinoshita:1995mt}
\begin{equation}
D_{\mu\nu}(q)=\left \{
  \begin{tabular}{cr}
  $\dfrac{i}{|\vec{q}|^{2}+\lambda^{2}+i\epsilon}$ & $\mu,\nu=0$ \\
  $\dfrac{i}{q^{2}-\lambda^{2}+i\epsilon}\bigg(\delta^{ij}-\dfrac{q^{i}q^{j}}{|\vec{q}|^{2}}\bigg)$ & $\mu=i \ne 0, \nu = j \ne 0$ \\
  0 & otherwise.  
  \end{tabular}
\right.
\end{equation}
\begin{itemize}
\item There are two Feynman rules for the photon propagator in Coulomb gauge.
\item[] The ``Coulomb" (timelike) propagator is
 $$
 \vcenter{\includegraphics[scale=1]{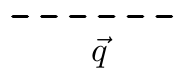}}
\hspace{-30em}  \frac{i}{\vec{q}^{\,\,2}+\lambda^2}
$$
\item[] The ``transverse" (spacelike) propagator is  
$$
 \vcenter{\includegraphics[scale=1]{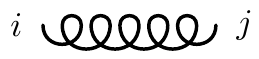}}
\hspace{-30em}  \dfrac{i}{q^{2}-\lambda^{2}+i\epsilon}\bigg(\delta^{ij}-\dfrac{q^{i}q^{j}}{\vec{q}^{\,\,2}-\lambda^{2}}\bigg)
$$
\end{itemize}
\subsection{Vertices}
The NRQED operators can be classified by  the power of $M$ by which they are suppressed. We need Feynman rules from operators suppressed by up to and including two powers of $M$.
\begin{itemize}
\item The operator $\psi^\dagger iD_t\psi$ gives rise to a one-photon  Feynman rule
$$
 \vcenter{\includegraphics[scale=0.75]{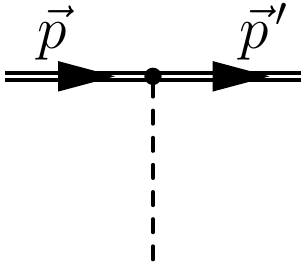}}
\hspace{-30em}  -iZe.
$$
\item The operator $\psi^\dagger \dfrac{\bm {D}^2}{2M}\psi$ gives rise to a one-photon Feynman rule
$$
 \vcenter{\includegraphics[scale=0.75]{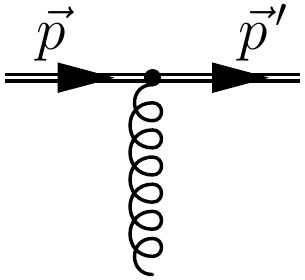}}
\hspace{-30em}  iZe\dfrac{\vec{p}+\vec p^{\,\prime}}{2M}
$$
and a two-photon Feynman rule 
$$
 \vcenter{\includegraphics[scale=0.75]{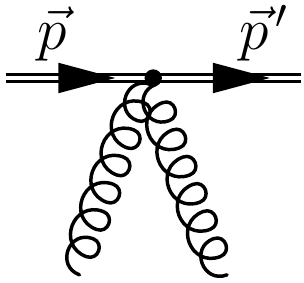}}
\hspace{-30em}   -\dfrac{Z^{2}e^{2}\delta^{ij}}{2M}.
$$
\item The operator $\psi^\dagger c_Fe\dfrac{\bm {\sigma\cdot B}}{2M}\psi$ gives rise to a one-photon Feynman rule 
$$
 \vcenter{\includegraphics[scale=0.75]{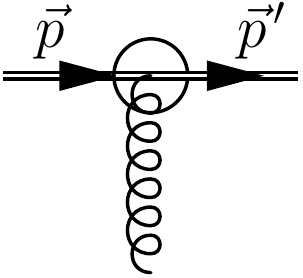}}
\hspace{-30em}   \dfrac{ec_{F}(\vec p^{\,\prime}-\vec{p})\times\vec{\sigma}}{2M}.
$$
\item The operator $\psi^\dagger c_De\dfrac{[\bm{\nabla\cdot E}]}{8M^2}\psi$ gives rise to a one-photon Feynman rule 
$$
 \vcenter{\includegraphics[scale=0.75]{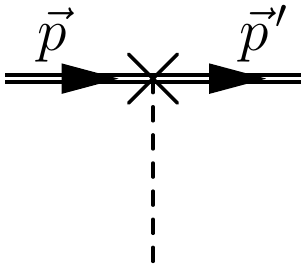}}
\hspace{-30em}   \dfrac{iec_{D}|\vec p^{\,\prime}-\vec{p}|^{2}}{8M^2}.
$$
In a non-Coulomb gauge, where $\nabla\cdot A\neq 0$, there is an additional one-photon Feynman rule 
$$
 \vcenter{\includegraphics[scale=0.75]{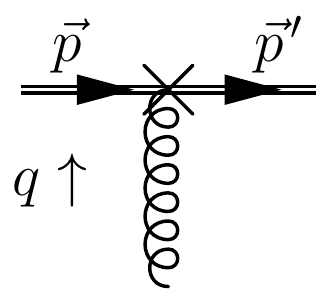}}
\hspace{-30em}   \dfrac{-iec_{D}q^0\vec{q}}{8M^2}.
$$
\item The operator $\psi^\dagger ic_Se\dfrac{\bm{\sigma}\cdot\left(\bm{D\times E}-\bm{E\times D}\right)}{8M^2}\psi$ gives rise to the one-photon Feynman rules 
$$
 \vcenter{\includegraphics[scale=0.75]{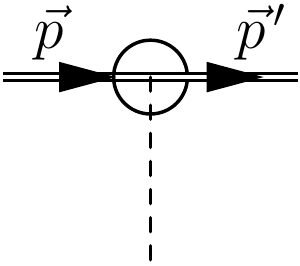}}
\hspace{-30em}  \dfrac{ec_{S}(\vec p^{\,\prime}\times\vec{p}\,)\cdot\vec{\sigma}}{4M^{2}},
$$
and
$$
 \vcenter{\includegraphics[scale=0.75]{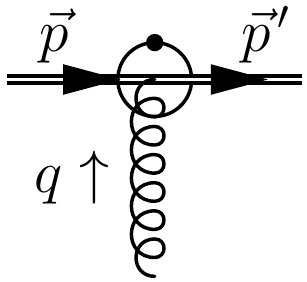}}
\hspace{-30em}   -\dfrac{ec_{S}q^{0}(\vec p^{\,\prime}+\vec{p})\times\vec{\sigma}}{8M^{2}},
$$
as well as the two-photon Feynman rules 
$$
 \vcenter{\includegraphics[scale=0.75]{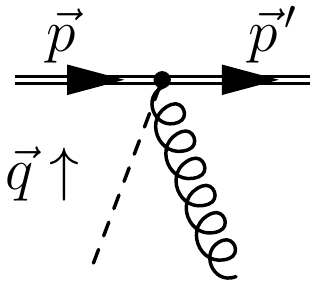}}
\hspace{-30em}   \dfrac{e^{2}c_{S}\,\vec{q}\times\vec{\sigma}}{4M^{2}},
$$
and
$$
 \vcenter{\includegraphics[scale=0.75]{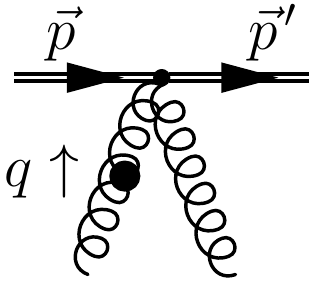}}
\hspace{-30em} -\dfrac{e^{2}c_{S}\sigma^{i}\epsilon^{ijk}q^{0}}{4M^{2}}.
$$
\end{itemize}

\end{document}